\documentclass[12pt,preprint,epsfig]{aastex}

\shorttitle{The VMC Survey. XII.}
\shortauthors{Chengyuan Li et al.}
\usepackage{color}

\begin{document}

\title{The VMC Survey. XI. Radial Stellar Population Gradients in the
  Galactic Globular Cluster 47 Tucanae}

\author{
Chengyuan Li\altaffilmark{1,2,3},
Richard de Grijs\altaffilmark{1,2},
Licai Deng\altaffilmark{3},
Stefano Rubele\altaffilmark{4},
Chuchu Wang\altaffilmark{2},
Kenji Bekki\altaffilmark{5},
Maria-Rosa L. Cioni\altaffilmark{6,7},
Gisella Clementini\altaffilmark{8},
Jim Emerson\altaffilmark{9},
Bi-Qing For\altaffilmark{5},
Leo Girardi\altaffilmark{4},
Martin A. T. Groenewegen\altaffilmark{10},
Roald Guandalini\altaffilmark{11},
Marco Gullieuszik\altaffilmark{4},
Marcella Marconi\altaffilmark{12},
Andr\'es E. Piatti\altaffilmark{13,14},
Vincenzo Ripepi\altaffilmark{12},
and
Jacco Th. van Loon\altaffilmark{15}
}

\altaffiltext{1} {Kavli Institute for Astronomy \& Astrophysics,
  Peking University, Yi He Yuan Lu 5, Hai Dian District, Beijing
  100871, China; joshuali@pku.edu.cn, grijs@pku.edu.cn}
\altaffiltext{2} {Department of Astronomy, Peking University, Yi He
  Yuan Lu 5, Hai Dian District, Beijing 100871, China}
\altaffiltext{3} {Key Laboratory for Optical Astronomy, National
  Astronomical Observatories, Chinese Academy of Sciences, 20A Datun
  Road, Chaoyang District, Beijing 100012, China} 
\altaffiltext{4} {INAF-Osservatorio Astronomico di Padova, vicolo
  dell'Osservatorio 5, I-35122 Padova, Italy} 
\altaffiltext{5} {ICRAR M468, The University of Western Australia, 35
  Stirling Highway, Crawley, WA, 6009, Australia} 
\altaffiltext{6} {Department of Physics, Astronomy, and Mathematics,
  University of Hertfordshire, Hatfield AL10 9AB, UK} 
\altaffiltext{7} {Leibnitz-Institut f\"ur Astrophysik Potsdam, An der
  Sternwarte 16, D-14482 Potsdam, Germany} 
\altaffiltext{8} {INAF-Osservatorio Astronomico di Bologna, Via
  Ranzani 1, I-40127 Bologna, Italy} 
\altaffiltext{9} {Astronomy Unit, School of Physics and Astronomy,
  Queen Mary University of London, Mile End Road, London E1 4NS, UK}
\altaffiltext{10} {Royal Observatory of Belgium, Ringlaan 3, 1180,
  Ukkel, Belgium} 
\altaffiltext{11} {Instituut voor Sterrenkunde, KU Leuven,
  Celestijnenlaan 200D 2401, 3001, Leuven, Belgium}
\altaffiltext{12} {INAF-Osservatorio Astronomico di Capodimonte, via
  Moiariello 16, 80131 Naples, Italy}
\altaffiltext{13} {Observatorio Astro\'nomico, Universidad Nacional de
  C\'ordoba, Laprida 854, 5000, C\'ordoba, Argentina}
\altaffiltext{14} {Consejo Nacional de Investigaciones
  Cient\'{\i}ficas y T\'ecnicas, Av. Rivadavia 1917, C1033AAJ, Buenos
  Aires, Argentina}
\altaffiltext{15} {Astrophysics Group, Lennard-Jones Laboratories,
  Keele University, ST5 5BG, UK}

\begin{abstract}
We present a deep near-infrared color--magnitude diagram of the
Galactic globular cluster 47 Tucanae, obtained with the Visible and
Infrared Survey Telescope for Astronomy (VISTA) as part of the VISTA
near-infrared $Y, J, K_{\rm s}$ survey of the Magellanic System
(VMC). The cluster stars comprising both the subgiant and red-giant
branches exhibit apparent, continuous variations in
color--magnitude space as a function of radius. Subgiant-branch stars
at larger radii are systematically brighter than their counterparts
closer to the cluster core; similarly, red-giant-branch stars in the
cluster's periphery are bluer than their more centrally located
cousins. The observations can very well be described by adopting an
age spread of $\sim 0.5$ Gyr as well as radial gradients in both
the cluster's helium abundance (Y) and metallicity ($Z$), which change
gradually from $({\rm Y} = 0.28, Z = 0.005)$ in the cluster core to
$({\rm Y} = 0.25, Z = 0.003)$ in its periphery. We conclude that the
cluster's inner regions host a significant fraction of
second-generation stars, which decreases with increasing radius; 
  the stellar population in the 47 Tuc periphery is well approximated
  by a simple stellar population.
\end{abstract}


\section{Introduction}

Stars in a given star cluster are usually assumed to have originated
from the same progenitor molecular cloud, from which they formed at
approximately the same time. In turn, this implies that they would
share the same metallicity (at least within a narrow range determined
by the metallicity of the progenitor molecular cloud). These
assumptions, which form the basis of the so-called (simple) single
stellar population (SSP) model, have led to numerous very successful
star cluster analyses. If we also assume that most stars evolve in
isolation, then one can use a unique theoretical isochrone to describe
the combination of main-sequence (MS), subgiant-branch (SGB), and
red-giant-branch (RGB) stars. However, for many intermediate-age star
clusters, and certainly for old globular clusters (GCs), the SSP
approximation seems to break down.

The discovery of striking, extended MS turnoffs (eMSTOs) in many
intermediate-age and old star clusters has led to a resurgence of the
field of stellar population synthesis. Most authors suggest that the
observed eMSTOs in, e.g., NGC 1783 \citep{Mack08,Rube13}, NGC 1846
\citep{Mack07,Rube13}, or NGC 1868 \citep{Li14}, can be adequately
described by assuming a cluster-internal age dispersion of roughly 300
Myr \citep{Rube10,Goud11,Mack13}, which hence challenges the
applicability of the SSP approach. Alternative models, which maintain
the SSP assumption, suggest that fast stellar rotation may be the
cause of the observed eMSTOs \citep{Bast09,Bast13,Li12,Li14,Yang13}
\citep[but see][]{Gira11,Plat12}. For some GCs, the breakdown of the
SSP model assumptions is particularly convincing. For instance, some
old GCs display double or multiple MSs, such as NGC 2808
\citep{Piot07} and NGC 6397 \citep{Milo12a}; in addition,
\cite{Milo08} found that NGC 1851 is characterized by two distinct
SGBs. Some GCs---such as $\omega$ Centauri \citep{Piot05,Soll07}, NGC
288 \citep{Piot13}, and M22 \citep{Lee09}---exhibit double or multiple
MSs, SGBs, and/or RGBs; Terzan 5 even shows clear, double HB clumps
\citep{Ferr09}. All of these observations strongly indicate that at
least some of the very old GCs contain multiple stellar populations.

The Galactic GC 47 Tucanae (47 Tuc; NGC 104) is another typical target
cluster that convincingly displays multiple stellar population
features across its entire color--magnitude diagram (CMD). It has also
been known for a long time that 47 Tuc displays a clear nitrogen (N)
dichotomy \citep{Norr79,Norr82,Norr84}. The cluster has been found to
host at least two distinct MSs \citep{Milo12b}, a broadening of the
SGB or (alternatively) multiple SGBs \citep{Ande09,Milo12b}, as well
as double or multiple RGBs \citep{Milo12b,Mone13}. In addition,
\cite{Nata11} analyzed the radial gradients in RGB-bump and
horizontal-branch (HB) stars, which indicate the presence of a
helium-enriched second stellar generation in 47 Tuc. \cite{Cord14}
determined a Na--O anticorrelation in 47 Tuc, a signature typically
seen in Galactic GCs \citep[e.g.,][]{Carr09} and which indicates
contamination by products of the proton-capture process
\citep{Lang93}. In other GCs \citep[e.g., in M71; see][]{Rami02} such
anticorrelations are even found in some less evolved MS stars, which
have not reached sufficiently high temperatures to trigger proton
capture. This implies the presence of more than one stellar population
in Galactic GCs.

Different origins have been proposed to account for the secondary
stellar generations; they mainly invoke scenarios involving accretion
of the ejecta of intermediate-mass stars \citep{Dant83,Renz83},
rapidly rotating massive stars \citep{Decr07}, massive binaries
\citep{Mink09}, and evolved asymptotic-giant-branch stars
\citep{Vent09}. Previous studies have shown that second-generation
stars originating from their first-generation counterparts are
generally more centrally concentrated
\citep{Decr08,Derc08,Bekk11}.\footnote{Full mixing of both populations
  will take at least 20 half-mass relaxation times
  \citep[cf.][]{Vesp13}. Note that the half-mass radius of 47 Tuc is
  $174''$ \citep{Trag93}, which implies that the vast majority of the
  cluster's member stars analyzed here are located outside of the
  cluster's half-mass radius and thus unrelaxed.} Indeed, in 47 Tuc
\cite{Milo12b} investigated the radial behavior of two distinct RGB
populations, as well as that of two subgroups of HB stars. They
concluded that second-generation stars gradually start to dominate at
increasingly smaller radii. This conclusion was confirmed by
\citet[][their figure 4]{Cord14} based on chemical abundance
analysis. \cite{Ande09} explored the global characteristics of the SGB
stars in 47 Tuc. They pointed out that the broadening of the SGB in
the 47 Tuc CMD indicates the presence of more than one stellar
population, but because of constraints inherent to their sample
selection, they could not investigate a promising radial trend related
to a possible second generation of SGB stars in 47 Tuc.

In this article, we present a deep, large-area, near-infrared (NIR)
CMD of 47 Tuc, obtained with the 4 m Visible and Infrared Survey
Telescope for Astronomy (VISTA). We confirm that the cluster's SGB and
RGB are significantly broadened, which cannot be explained as simply
owing to the effects of either photometric uncertainties or
(differential) extinction. Compared with \cite{Milo12b}, the RGB's
radial behavior we find displays a more obvious radial color bias. We
find that the average color of the RGB stars at large radii is
significantly different from that in the innermost regions. The
apparent color gradient strongly indicates that the two RGB
populations contain significantly different fractions of stars formed
as part of a second stellar generation. Following \cite{Ande09}, we
also find that the cluster's SGB is significantly broadened. The SGB
stars in the periphery are on average significantly brighter than the
innermost SGB stars. All of these features are so striking that they
can even be clearly seen based on a single, quick glance at the
cluster's global CMD. The present paper focuses on new, high-quality
NIR observations of 47 Tuc, which offers an excellent comparison data
set to the ultraviolet/optical study of \cite{Milo12b}. We also
highlight that our observations cover a significantly larger field
than that of \cite{Milo12b}: our data set's maximum distance to the
cluster's central region reaches 4000$''$ (compared with the radial
range out to 1500$''$ available in the earlier
publication).\footnote{At the cluster's distance of 4.6 kpc (derived
  below), $1'' \equiv 0.022$ pc.} We can thus very well characterize
the effects caused by background contamination.

This article is organized as follows. In Section 2 we describe the
data reduction approach. Section 3 presents the main results, which we
discuss in Section 4. A brief summary and our conclusions are
presented in Section 5.

\section{Data Reduction and Analysis}

The data set analyzed in this article was obtained as part of the {\it
  VISTA near-infrared $Y, J, K_{\rm s}$ survey of the Magellanic
  System} (VMC;\footnote{http://star.herts.ac.uk/$\sim$mcioni/vmc} PI
M.-R. L. Cioni: see Cioni et al. 2011). The survey started in November
2009 and is expected to extend beyond the originally planned 5 yr time
span \citep[survey completion is foreseen for 2018;][]{Arna13}. Its
main goals and first data are described in \cite{Cioni11}. 47 Tuc is
fortuitously located in front of the Small Magellanic Cloud (SMC), on
VMC tile ``SMC 5$_{-}$2,'' which made it an ideal target for early VMC
data acquisition.

The VMC images were reduced by the Cambridge Astronomy Survey Unit
(CASU) using the VISTA Data Flow System (VDFS) pipeline
\citep{Irwi04}. To perform our point-spread-function (PSF) photometry
starting from the pawprint VMC images, we selected---in all three
filters ($Y,J,K_{\rm s}$)---epochs with limited seeing
inhomogeneities. The selected reduced data, comprising two epochs in
$Y$ and $J$ and nine in $K_{\rm s}$, are PSF-homogenized and stacked
to obtain a deep tile image (using a novel method that will be
described in detail by Rubele et al., in prep.). We performed PSF
photometry on the deep image of tile SMC 5$_{-}2$, using the {\sc
  iraf} {\sc daophot} package \citep{Stet87}. The {\sc psf} and {\sc
  allstar} tasks were used to produce PSF models and perform the final
photometry. We checked and corrected our PSF photometry for aperture
effects using catalogs retrieved from the VISTA Science Archive
\citep{Cros12} for the bulk of the observed stars.

Here, we focus our analysis on the $(Y, Y-K_{\rm s})$ CMD, where the
features of interest are most apparent, because this color extends
over the longest available wavelength range. However, we confirmed
that our conclusions also hold based on analysis of the $(Y,Y-J)$ and
$(J,J-K_{\rm s})$ CMDs. Our observations cover an area of roughly
$1.2\times5$ deg$^2$. The bulk of the 47 Tuc stars are located toward
the southwest of this region, so that the opposite corner of the field
is suitable for analysis of the field-star properties, which we will
use to decontaminate the cluster CMD.

47 Tuc is an extremely large and crowded GC. In our data, it covers
more than 300,000 stars with $Y \in [11.0,25.5]$ mag. Because of the
effects of mass segregation \citep[cf.][]{grijs02a,grijs02b,grijs02c},
almost all of the more massive ($m_\ast\geq0.88M_{\odot}$), bright
($Y\leq13$ mag) cluster stars are contained within the central region,
which dramatically increases the background level in the cluster core
because of the extended, faint wings of the PSF profile. In the
central region, almost all stars fainter than $Y \sim 13$--15 mag will
fade into the background because of a combination of the enhanced
background level and stellar crowding in our ground-based images. For
these reasons, we only use the 3071 stars brighter than $Y = 13.0$
mag, in essence HB stars and a fraction of the cluster's bright RGB
and asymptotic-giant-branch stars, to determine the cluster
center. Our approach to find the cluster center is identical to that
employed by \cite{grijs13} and \cite{Li13,Li14}.

We first selected a sufficiently large rectangular area that clearly
contained the full cluster region and then divided the stellar spatial
distributions into 500 bins in both the right ascension ($\alpha_{\rm
  J2000}$) and declination ($\delta_{\rm J2000}$) directions. In both
spatial coordinates, the stellar number densities follow Gaussian-like
profiles, which allow us to determine the coordinates corresponding to
the two-dimensional (2D) maximum stellar density. The resulting center
coordinates are $\alpha_{\rm J2000} =00^{\rm h}24^{\rm m}04.80^{\rm s}
(6.020^{\circ}$; statistical uncertainty $0.007^{\circ} \equiv 25''),
\delta_{\rm J2000} = -72^{\circ}04'48'' (-72.080^{\circ} \pm
0.006^{\circ} \equiv 20')$. Our approach introduces negligible biases,
because the member stars of extremely old GCs should all have adapted
their kinematics to be in equilibrium with the prevailing
gravitational potential, so that both the bright stars (which we
excluded from our analysis) and their fainter companions experience
the same gravitational potential. Our result is consistent with
previous determinations of the cluster center by \cite{Mcla06},
\cite{Gold10}, and \cite{Harr10}.

We subsequently selected a second region centered at $\alpha_{\rm
  J2000} =00^{\rm h}31^{\rm m}59.00^{\rm s} (8.00^{\circ})$,
$\delta_{\rm J2000} = -71^{\circ}42'00'' (-71.70^{\circ})$, with a
radius of 600$''$, to calculate the background stellar number density
(for stars with $Y\leq 13.0$ mag).  The vast majority of stars in this
region are foreground Milky Way and background SMC stars;
contamination by 47 Tuc members at these large radii is $\ll 15$\%, an
estimate based on careful assessment of the cluster's extended radial
profile. This region is a suitable choice for a ``field'' region,
because its center is located more than 2400$''$ from the cluster
center, a distance that is close to the cluster's tidal radius
\citep[$r_{\rm t} = 56$ pc $\equiv 2500''$ for a distance modulus of
  $(m-M)_0 = 13.40$ mag or a distance of 4.6 kpc, derived below;
  see][]{Harr96,Lane12}; this is unlikely to significantly affect our
field-star characterization. We cannot rule out the presence of a
significant population of extratidal stars, however. 
  \cite{Lane12} predicted that 47 Tuc should exhibit at least one
  extratidal tail, which would cross our observational field of view
  from the northwest to the southeast. We hence selected a region in
the northeast of our field of view as field reference so as to avoid
contamination by extratidal-tail stars as much as possible. In fact,
we found that the stellar number density in the southeast is indeed
significantly higher than that in the northeast. We next compared the
cluster's stellar number density with that of the background field to
check at which radius, $R_{\rm f}$, the cluster's stellar number
density becomes indistinguishable from the background level. We
determined $R_{\rm f} = 1220\pm 20''$, and hence adopted a cluster
``size'' of 1200$''$ for further study, equivalent to approximately
50\% of its tidal radius. This is a conservative choice, because we
eliminated the brightest stars from our analysis. Mass segregation
will cause an underestimation of the ``bright'' cluster size compared
with that resulting from having adopted a homogeneous stellar
sample. However, even if we were to select a smaller cluster size, our
results still strongly indicate the presence of multiple stellar
generations.

\begin{figure}[ht!]
\plotone{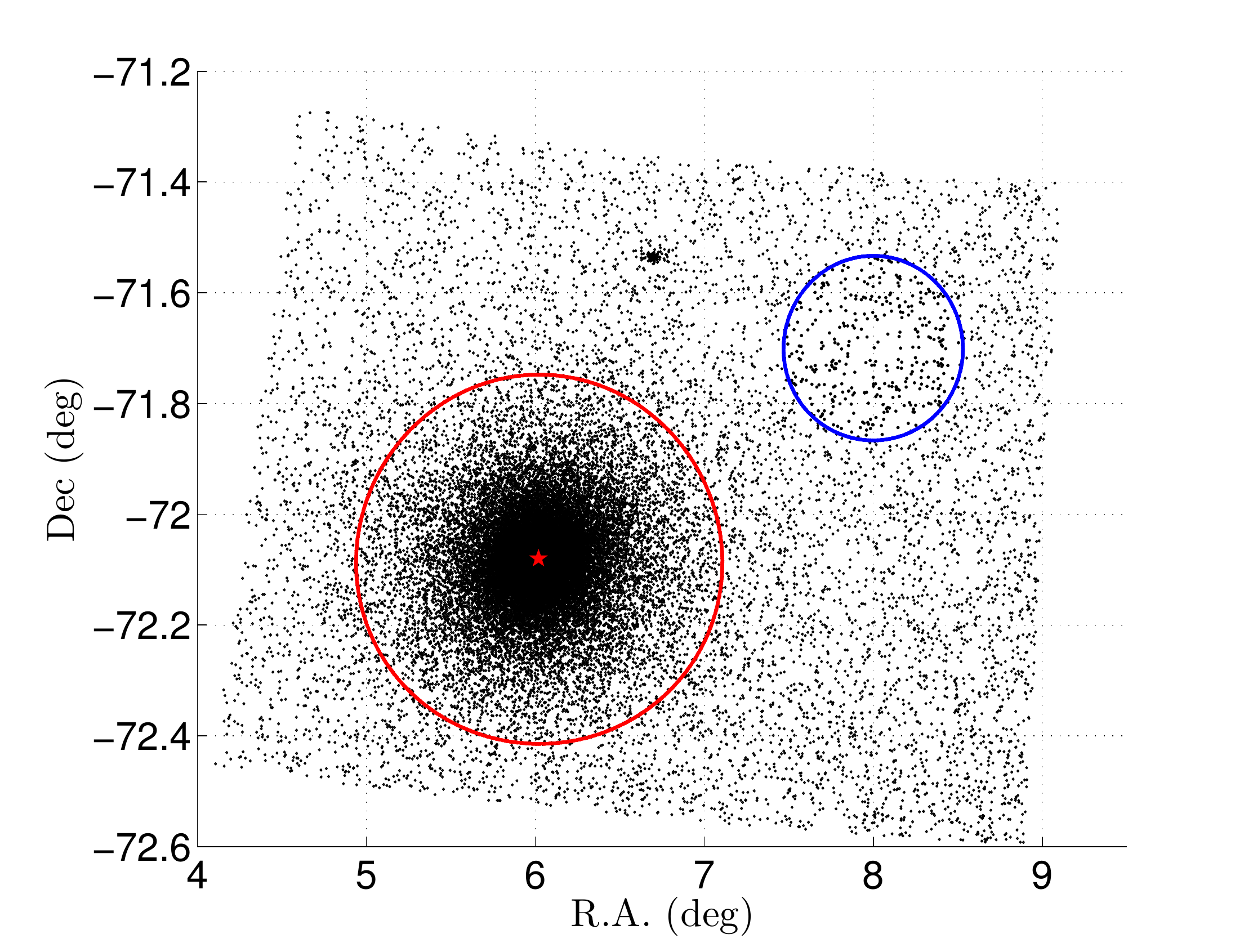}
\caption{Spatial distribution of 47 Tuc stars with $Y\leq 17.0$
  mag. North is up; East is to the left. The red pentagram represents
  the adopted cluster center, the red circle indicates the cluster
  region of interest, corresponding to a radius of 1200$''$, and the
  blue circle maps out a representative region adopted for field-star
  decontamination, with a radius of 600$''$. The small group of stars
  located at approximately ($\alpha_{\rm J2000}\sim6.7^\circ,
  \delta_{\rm J2000}\sim-71.5^\circ$) is the SMC cluster NGC 121. The
  center of the SMC is located at approximately $1.7^\circ$ toward the
  southeast of 47 Tuc.}
\label{F1}
\end{figure}

\begin{figure}[ht!]
\plotone{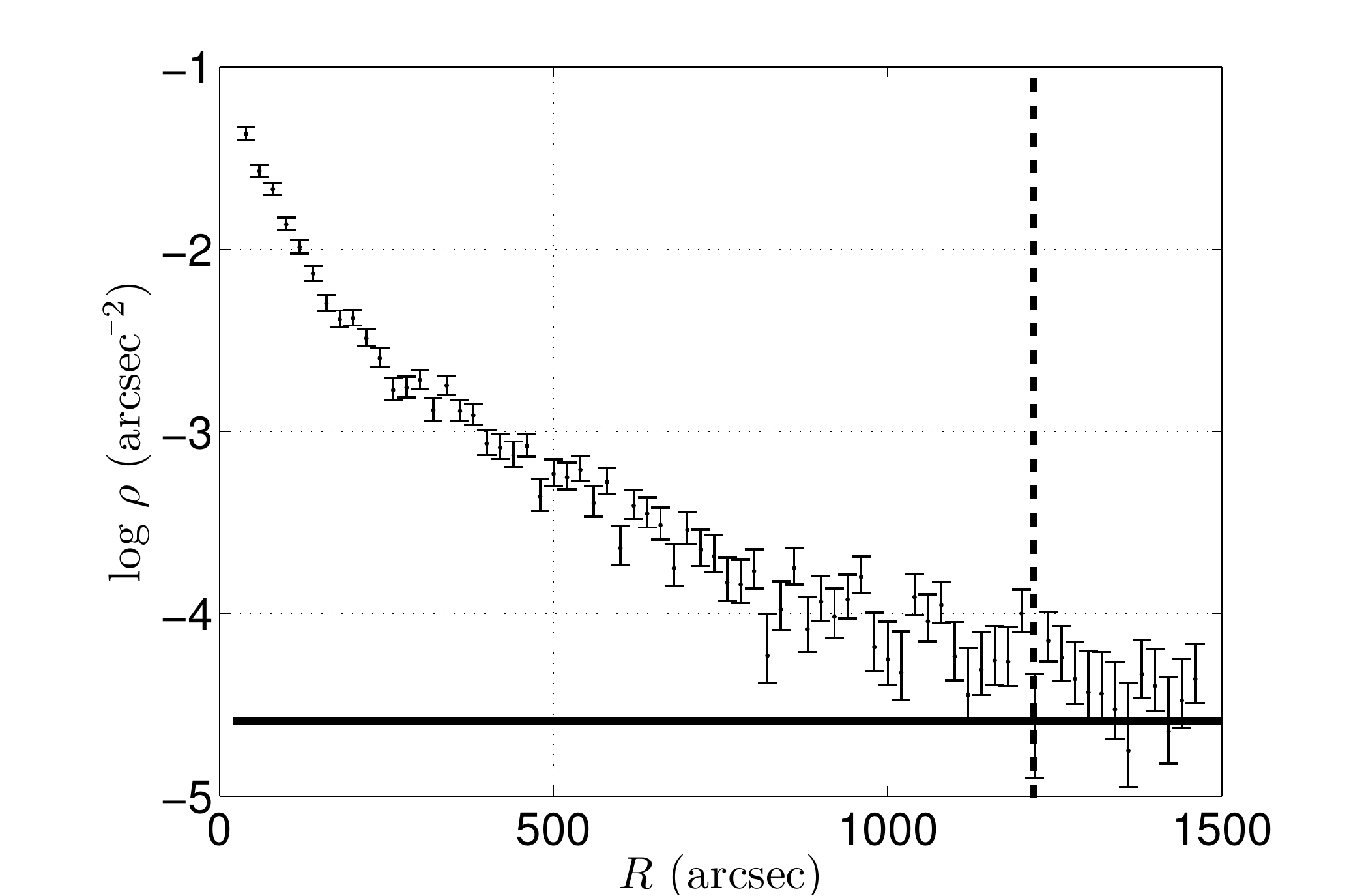}
\caption{Logarithmic radial number-density profile of stars brighter
  than $Y = 13$ mag. The horizontal solid line indicates the
  background level, while the vertical dashed line signifies the
  adopted cluster size.}
\label{F2}
\end{figure}

Figure \ref{F1} displays the stellar spatial distribution of stars
brighter than $Y = 17$ mag. We also show the stars in the adopted
field region (blue circle), characterized by a radius of 600$''$. The
cluster region is indicated by the red circle. In Figure \ref{F2} we
display the radial stellar number-density profile (on a logarithmic
scale), where the black solid line indicates the field stellar density
level and the vertical black dashed line indicates the radius, $R_{\rm
  f}$, where the cluster's stellar number density becomes
indistinguishable from that of the background.

Next, we adopted the most suitable isochrone describing our
photometric data, using model isochrones based on the {\sc pgpuc}
stellar evolution
code\footnote{http://www2.astro.puc.cl/pgpuc/iso.php} \citep[][with
  $Z_\odot = 0.0152$ as its basis]{Valc12}. We adopted an age for 47
Tuc of 12.5 Gyr or $\log(t \mbox{ yr}^{-1})=10.10$
\citep[cf.][]{Zocc01,Mcla06,DiCr10,Mcdo11} and a metallicity of
$Z=0.0042$ or ${\rm [Fe/H]}=-0.55$ dex. The latter value is consistent
with the determination of \cite{Nata11} (Note that these authors
adopted [M/H] $= -0.52$ dex instead of [Fe/H]; [M/H] = $A \times$
[Fe/H], where $A \in [0.9,1.0]$. We simply adopt an average value of
$A = 0.95$.) The adopted extinction, $E(B - V)=0.04$ mag, is equal to
that determined and adopted by \cite{Harr96}, \cite{Ande09}, and
\cite{Cord14}. We also need to adopt a suitable value for the
[$\alpha$/Fe] ratio. \cite{Carr09} determined [$\alpha$/Fe] = 0.40
dex. We hence adopted the maximum value available in the {\sc pgpuc}
code, [$\alpha$/Fe] = 0.30 dex, to generate the most representative
isochrone for the bulk of the cluster stars; we found that this small
offset between the most appropriate and closest-available
model-[$\alpha$/Fe] values will not introduce significant fitting
problems. Based on these input parameters, the best-fitting distance
modulus to 47 Tuc is $(m - M)_0= 13.40\pm0.10$ mag, which is
consistent with the determinations of \cite{Harr96}, \cite{Zocc01},
and \cite{Mcdo11}.

We next performed field-star decontamination. As illustrated in Figure
\ref{F1}, we selected stars from a suitably chosen nearby region as
our reference to statistically correct for background
contamination. We selected a field region covering a circular area
with a radius of 600$''$, which corresponds to one quarter of the
adopted cluster area, and compared the corresponding CMDs. For every
star in the CMD of the field region, we removed the closest four
counterparts (thus correcting for the difference in the cluster versus
field areas adopted) in the corresponding CMD of the cluster. Although
any such star may not individually be a genuine field star, this
approach nevertheless ensures that the cluster CMD is decontaminated
statistically robustly \citep[e.g.,][]{grijs13,Li13,Li14}. In
addition, we mainly focus on SGB and RGB stars, while very few of the
field stars in the region adopted for field-star decontamination are
this bright. This hence supports the reliability of our statistical
decontamination approach.

\begin{figure}[ht!]
\plotone{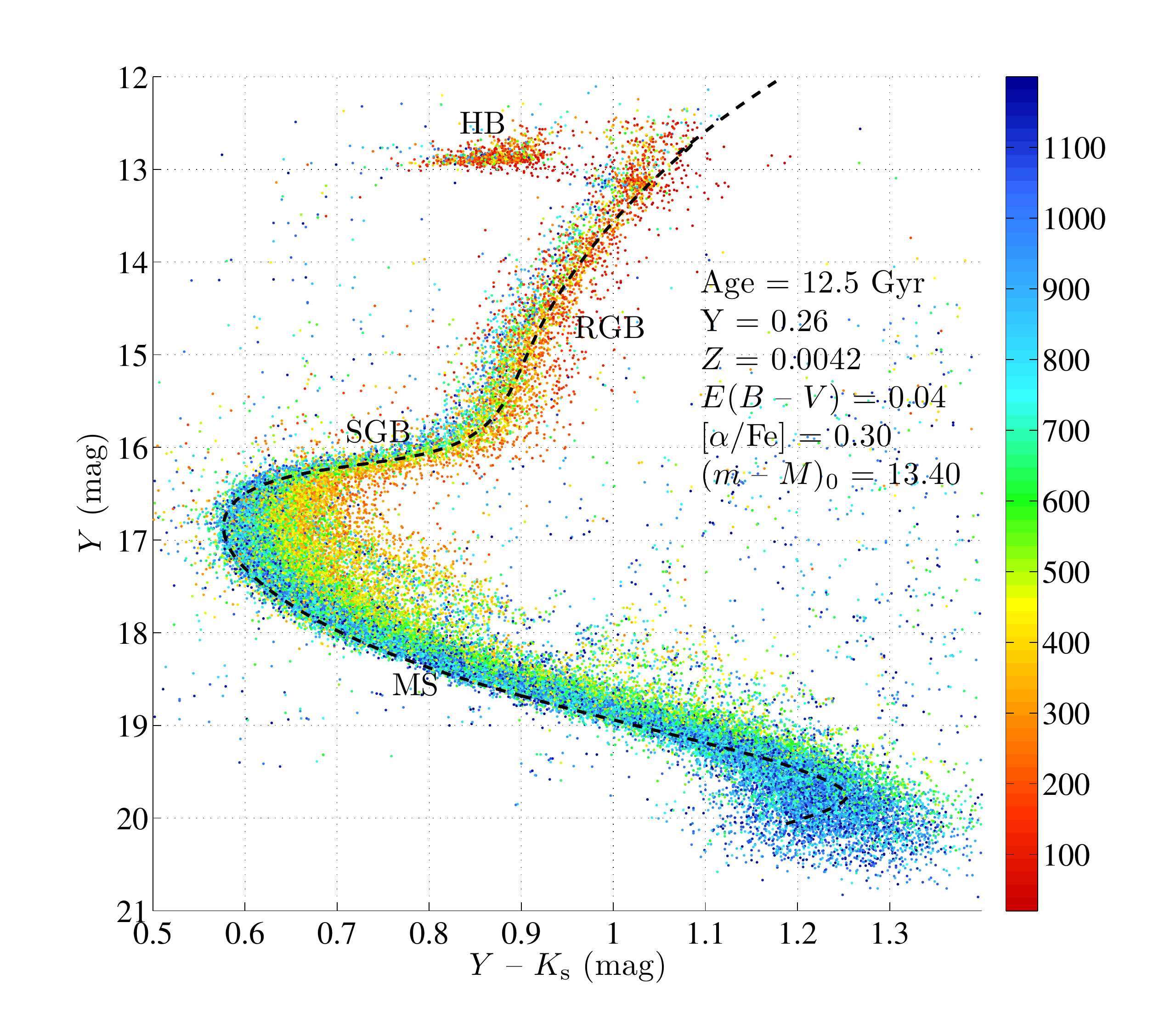}
\caption{$(Y,Y-K_{\rm s})$ CMD of 47 Tuc, showing only stars with
  photometric uncertainties of $\leq0.02$ mag in all filters. The
  color scale indicates the distance to the cluster center (in
  arcsec). The black dashed line represents the best-fitting
  isochrone.}
\label{F3}
\end{figure}

\begin{figure}[ht!]
\plotone{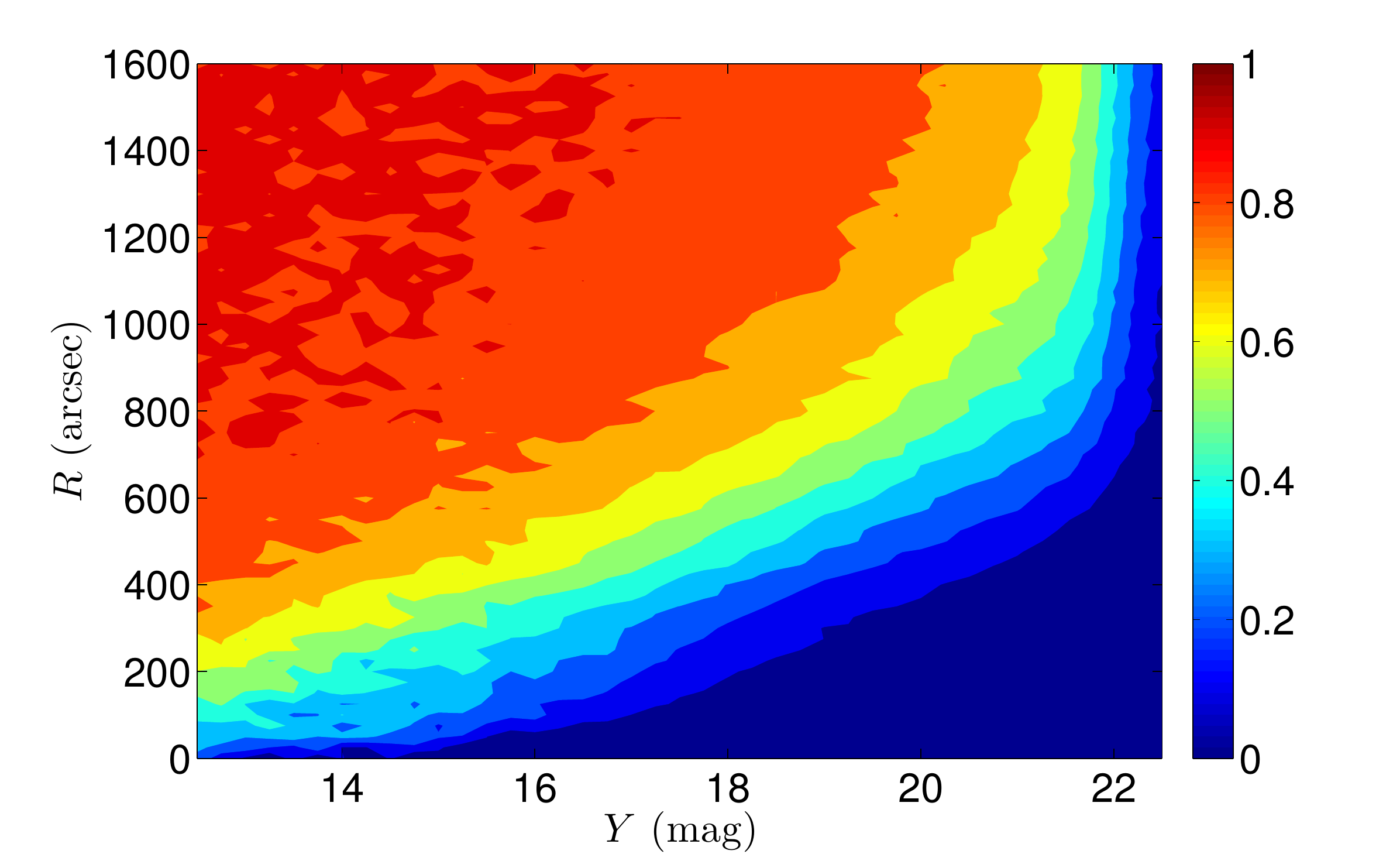}
\caption{2D completeness map as a function of radius and $Y$
  magnitude.}
\label{F4}
\end{figure}

Figure \ref{F3} shows the decontaminated cluster CMD, including all
stars located within 1200$''$ of the cluster center, as well as the
most appropriate isochrone given the cluster's overall parameters. The
color scale represents the distance to the cluster center. A quick
glance at this figure already suggests a systematic difference between
the cluster stars in the innermost region and in the cluster's
periphery. Even if we ignore the MS stars due to a lack of accurate
data in the cluster's core region (because of low completeness at
these relatively faint magnitudes in the inner region; see Figure
\ref{F4}), the SGB and RGB stars exhibit a clear difference as a
function of radius. SGB stars at large radii are systematically
brighter than those located in the central regions; they join the RGB
at clearly bluer colors compared with the color range characteristic
of the innermost RGB stars, hence implying higher characteristic
temperatures. All stars shown in Figure \ref{F3} have photometric
uncertainties of less than 0.02 mag. It is thus clear that the
broadening of both the SGB and the RGB cannot be only owing to
photometric uncertainties.

Because of the complicated physical conditions affecting our 47 Tuc
observations (including those caused by significant mass segregation
and frequent stellar blending due to crowding effects), most faint
stars have very large photometric uncertainties, while some of the
brightest stars ($K_{\rm s} \la 11.5$ mag) are partially
saturated. This latter effect also introduces large photometric
uncertainties. We therefore removed those stars in our overall sample
that have photometric uncertainties in excess of 0.02 mag in all three
filters. This resulted in a final working sample that contained the
$\sim$30\% of stars with the highest-accuracy photometry in the
corresponding magnitude range. Note that because of the crowding in
the cluster's core region, more centrally located stars will have
larger errors (for the same source brightness), which hence may
introduce an additional bias. To quantify such a bias, we used
artificial-star (AS) tests to study the systematic effects of crowding
on the uncertainties in the resulting PSF photometry. We added $7.2
\times 10^6$ ASs to the raw image and subsequently measured them in
the same manner as our sample of real stars, using PSF photometry. The
number of ASs is roughly 20 times larger than the number of real
stars. To avoid a situation in which the ASs affect or even dominate
the background level and cause crowding artifacts, we only add 3500
stars to the raw image at any given time. Although this choice
  implies that there is a (very small) probability that two ASs may
  blend with each other, adopting a much smaller number of ASs at a
  time would significantly increase the computational cost of our AS
  tests. (We quantified the effects of ASs blending with other ASs,
  and found the possibility of such a situation occurring to be
  negligible.) We repeated this procedure 2000 times. The
distribution of the ASs followed a spatial distribution that
reproduced the cluster shape derived from bright stars, as well as a
distribution in the color--magnitude diagram that resembled the real
distribution. The resulting AS catalog contains the input and recovered (output)
magnitudes, as well as the photometric errors computed as output - input
magnitudes.

We carefully assessed any systematic effects of the photometric
  uncertainties on our PSF photometry owing to crowding.  The main
  result of these AS tests is that the positions in the CMD of the SGB
  and RGB stars that we are concerned with in this study are
  negligibly affected. However, the photometric errors for stars in
  the central region are larger than the errors resulting from the use
  of {\sc daophot}, for the same object, while there is good agreement
  between both sets of photometric-error measurements in the outer
  regions, as expected. This effect only becomes important for stars
  that are fainter than the MSTO; for their brighter counterparts the
  deviation is negligible. Our results hence confirm that the spread
  in the SGB/RGB is real.

The AS tests also allow us to estimate the photometric completeness of
our stellar catalog. The crowding in the cluster's central regions
affects the ASs in a similar way as the real cluster members. We
generated artificial stars that were homogeneously distributed across
the same region as the observations. Those ASs that returned either no
photometric measurement or that were characterized by a difference in
output - input magnitude that was greater than five times their
photometric uncertainties were considered stars that could not be
recovered (hence tracing the level of completeness of the
observations). Figure \ref{F4} displays the cumulative 2D completeness
map as a function of both radius and $Y$ magnitude.

\section{Results}
\subsection{Subgiant-branch stars}

We start by focusing on the region occupied by the SGB stars. First,
we select a box delineated by $Y\in[15.5,16.7]$ mag and $(Y - K_{\rm
  s})\in[0.70,0.85]$ mag. Although this selection is quite arbitrary,
we only need to ensure that the region contains the vast majority of
SGB stars. Because the photometry of stars in this magnitude range is
more accurate than that of the faint(er) MS stars, we further impose a
maximum photometric uncertainty of 0.015 mag. This constraint leaves
us with the $\sim 30$\% of stars with the highest-accuracy photometry
in the corresponding part of CMD space.

\begin{figure}[ht!]
\plotone{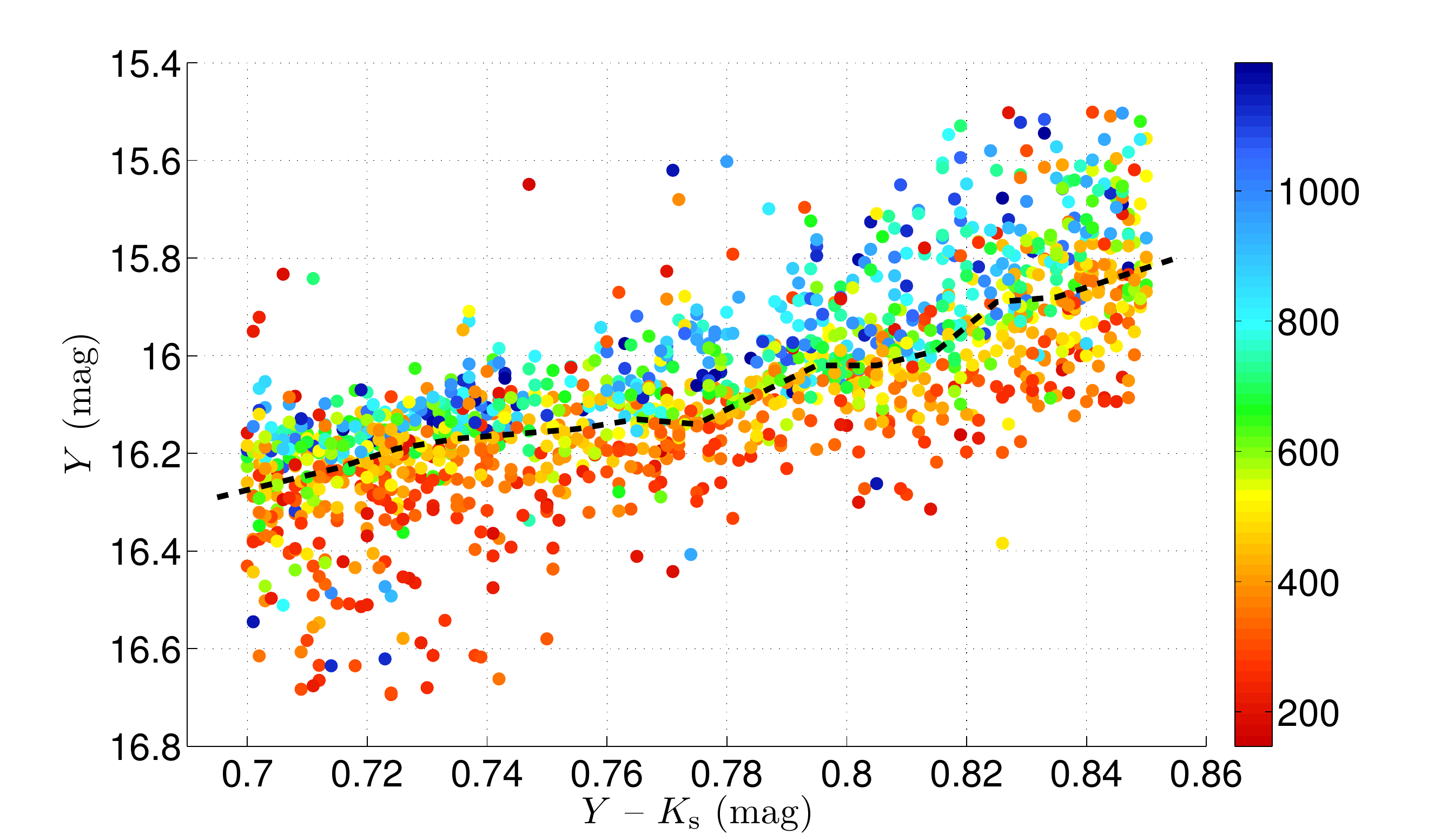}
\caption{CMD of the adopted SGB region. The color scale indicates the
  distance to the cluster center (in arcsec). Because of sampling
  incompleteness in the cluster core, the minimum radius attainable is
  $\sim 150''$. All stars have magnitudes within 3$\sigma_{\Delta Y} =
  0.54$ mag of the SGB ridge line.}
\label{F5}
\end{figure}

We next adopt the cluster-wide ridge line as our standard model. The
locus of the SGB ridge line is determined by connecting the loci of
the maximum stellar number density as a function of color. We
calculate the magnitude difference with respect to this observational
ridge line for all stars in our SGB sample, $\Delta Y = Y_{\rm
  SGB}-Y_{\rm iso}$. The resulting magnitude dispersion is distributed
in a Gaussian-like fashion with $\sigma_{\Delta Y}\sim0.18$ mag. We
remove stars that were found beyond 3$\sigma_{\Delta Y} = 0.54$ mag of
the ridge line, resulting in a sample of 1389 stars: see Figure
\ref{F5}.

\begin{figure}[ht!]
\plotone{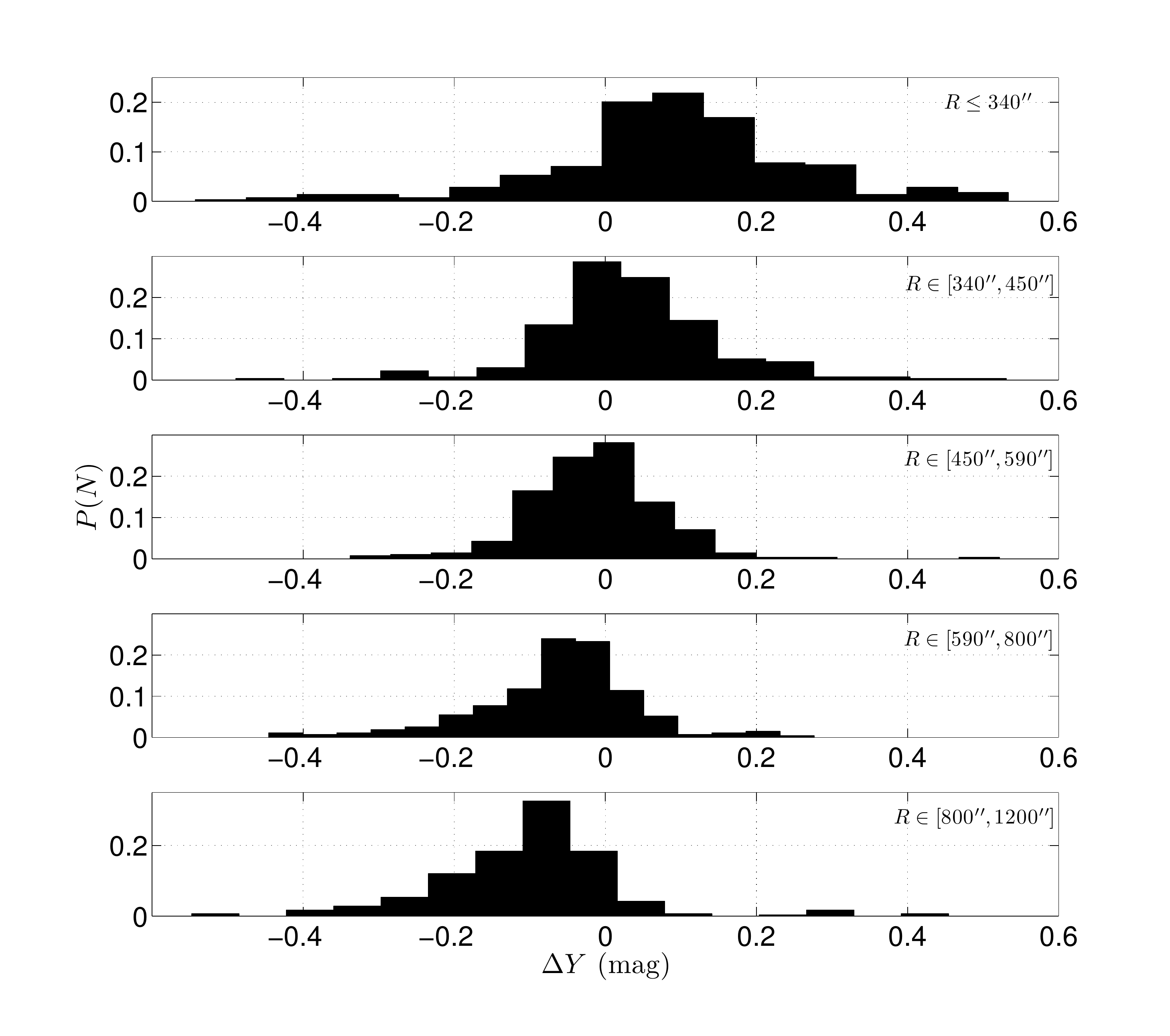}
\caption{Probability distribution of $\Delta Y$ as a function of
  radius for SGB stars in 47 Tuc. From top to bottom, the panels
  represent radial bins from the cluster's inner regions to its
  outskirts.}
\label{F6}
\end{figure}

Figure \ref{F5} shows that the SGB stars in the cluster's periphery
are systematically found above the ridge line, i.e., they are brighter
compared with their more centrally located counterparts. We divide the
SGB sample into five annuli as a function of radius to check whether
and---if so---how their magnitude dispersion varies with radius. The
four boundaries between subsequent annuli were set at $340''$,
$450''$, $590''$, and $800''$, which resulted in roughly equal stellar
numbers (ranging from 269 to 283) per radial bin. For each radial bin,
we calculate the distribution of the stellar magnitude dispersion,
normalized to the total number of stars in the relevant bin. We use
these distributions to calculate the probability distribution of
$\Delta Y$ in each radial bin, as shown in Figure \ref{F6}.

Based on Figure \ref{F6}, it is clear that the SGB stars in the
innermost radial bin (top panel) are systematically fainter ($\Delta Y
>0$ mag) than those in the cluster's periphery (bottom panel). As a
function of increasing radius, the mean magnitude of the bulk of the
SGB stars becomes gradually brighter ($\Delta Y<0$ mag). We also note
that for the central sample, the magnitude dispersion of the central
SGB stars is larger than that of the peripheral SGB sample. This is as
expected, because the observed stellar distribution is a 2D projection
onto the plane of the sky: the stars in the peripheral sample are
indeed located far from the cluster center, but the innermost stars
are significantly contaminated by stars that are physically located at
larger radii. Therefore, if the properties of the SGB stars at large
radii differ systematically from those in the innermost sample, the
contaminated innermost sample will exhibit a larger magnitude
dispersion.

\begin{figure}[ht!]
\plotone{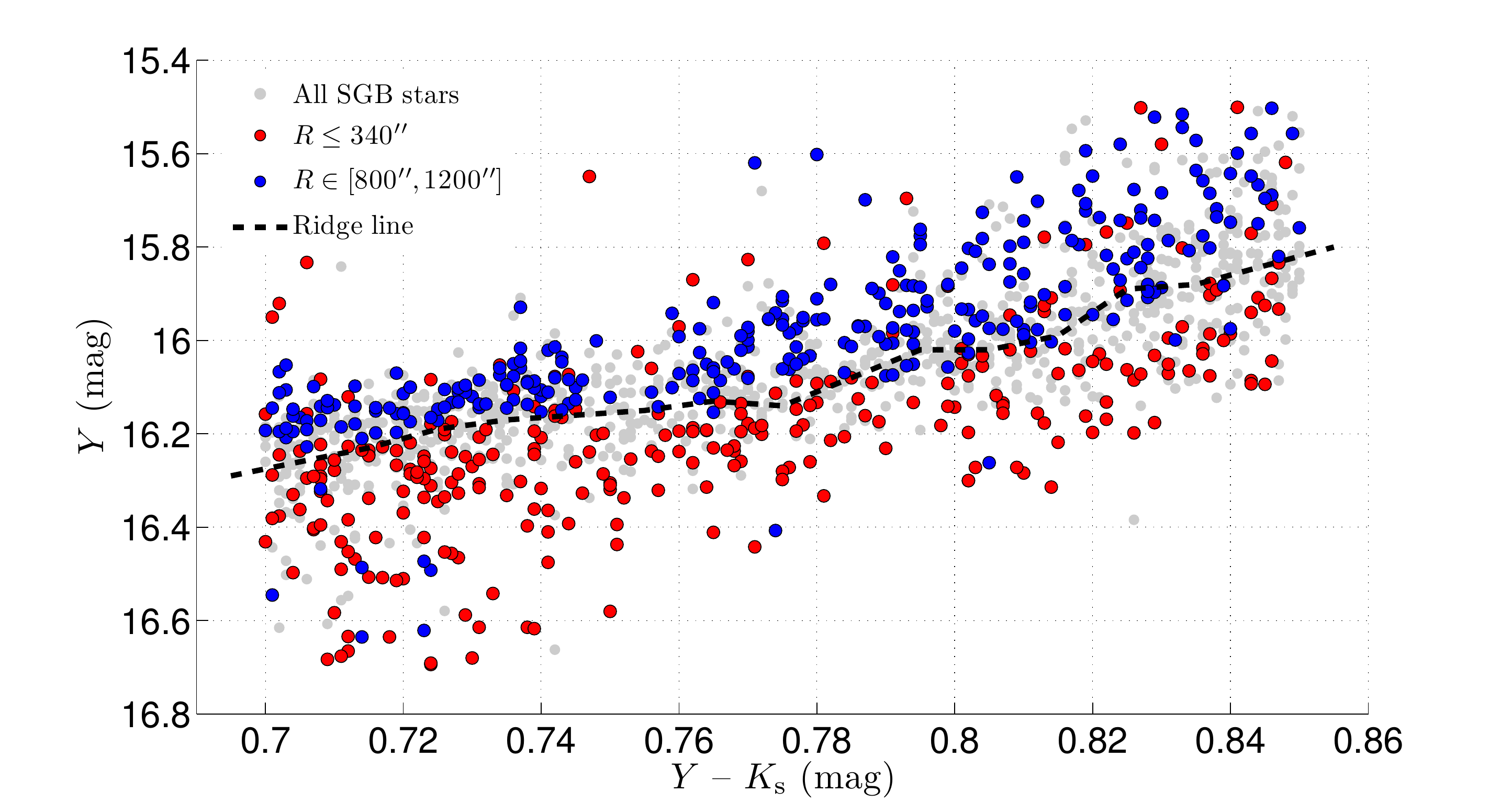}
\caption{CMD of the most centrally located SGB stars in 47 Tuc
  ($R\leq340''$: red solid bullets) and the outermost SGB stars
  ($R\in[800'',1200'']$: blue solid bullets). Smaller grey bullets
  represent all SGB stars in our sample.}
\label{F7}
\end{figure}

In addition, blending of unresolved stars that are physically not
associated with each other but located along the same line of sight
will also cause a dispersion of SGB stars to brighter magnitudes. Such
blending events can happen anywhere, but they are expected to occur
more frequently in the cluster's innermost regions, where the stellar
number density is highest. Both of these effects will cause a bias to
brighter magnitudes that predominantly affects the innermost SGB
stars. It is thus natural to expect that the intrinsic distribution of
SGB stars in the cluster core would be narrower than observed. In
Figure \ref{F7} we show the CMD of SGB stars at $R\leq340''$ (i.e.,
the innermost radial bin), as well as that composed of SGB stars at
$R\in[800'',1200'']$ (i.e., the outermost radial bin adopted). It is
clear that the most centrally located SGB stars are more dispersed (in
magnitude) than the peripheral SGB stars; both samples are clearly
different.

We use a Monte Carlo method to estimate how many projected stars at
large radii will contaminate the stellar sample drawn from the inner
regions \citep[cf.][]{Li14}. For example, if we want to estimate the
contamination of SGB stars in the ring defined by $R\in[R_1,R_2]$,
where $R_2>R_1$, we first calculate how many SGB stars are located at
$R \ge R_2$, say $N=N(R\geq{R_2})$. We then assume that these $N$
stars are located in a three-dimensional (3D) spherical shell at
$R\in[R_2,R_{\rm f}]$, where $R_{\rm f} = 1200''$ is the cluster
size. The 3D stellar density at $R\in[R_2,R_{\rm f}]$ is then
\begin{equation}
  \rho=\frac{3N}{4\pi(R_{\rm f}^3-R_2^3)}.
\end{equation}
We next generate a 3D spherical cluster of size $R_{\rm f}$, which
hence contains $N_{\rm art}=(4\pi{R_{\rm f}^3}/3)\times{\rho}$
artificial stars. The artificial stars are homogeneously distributed
with density $\rho$. We now randomly select a given direction to
represent the line-of-sight direction; the number of stars located at
$R\geq{R_2}$ that will be projected at $R_{\rm proj}\in[R_1,R_2]$ is
referred to as $N_{\rm proj}$. For each radial bin, we repeat this
process 10 times to calculate the average $N_{\rm proj}$ and obtain an
approximate projected number of stars in the radial annulus of
interest. Because only stars at $R\geq{R_2}$ will cause contamination
of a ring at $R\in[R_1,R_2]$, even if we assume that the average
density of stars at $R\geq{R_2}$ is homogeneous and that they are
spherically distributed, which will underestimate the real stellar
density in the region at $R\leq{R_2}$, this will not affect our
estimation.

The resulting ratio of the projected contamination indicates, for
example, that for the sample drawn from the (projected) innermost
radii, $R\leq340''$, 103 stars are actually physically located at
$R>340''$. Similarly, for the radial range $R\in[590'',800'']$, only
43 stars will be located beyond $R=800''$ along the line-of-sight
direction.

If we normalize the total number of stars that we observe in each
projected radial bin, the contamination due to line-of-sight
projection decreases from 36\% for the innermost radii ($R\leq340''$)
to 16\% for the penultimate radial range ($R\in[590'',800'']$). We
have assumed that for radii $R\in[800'',1200'']$ the effects of
projection are negligible. Because more than half of the central SGB
stars are, in fact, peripheral stars that have been projected along
the line-of-sight, it is not strange that the magnitude dispersion of
the innermost SGB stars is larger than that of the peripheral
sample(s). Table 1 includes the details of the projected contamination
fraction derived this way, for both SGB and RGB stars (for a
discussion of the latter, see Section 3.2).

\begin{table}[h!]
 \centering
 \begin{minipage}{110mm}
  \caption{Contamination due to line-of-sight projection for SGB and
    RGB stars as a function of radius.}
  \begin{tabular}{@{}l|c||l|c@{}}
  \hline
  \hline
SGB stars           & $f_{\rm proj}$ & RGB stars   & $f_{\rm proj}$ \\ 
\hline\hline
$R\leq340''$        & 36.40\% & $R\leq260''$        & 29.00\% \\
$R\in[340'',450'']$ & 20.82\% & $R\in[260'',370'']$ & 20.55\% \\
$R\in[450'',590'']$ & 20.42\% & $R\in[370'',510'']$ & 19.46\% \\
$R\in[590'',800'']$ & 15.87\% & $R\in[510'',730'']$ & 16.26\% \\
\hline\hline
\end{tabular}
\end{minipage}
\end{table}

\begin{figure}[ht!]
\plotone{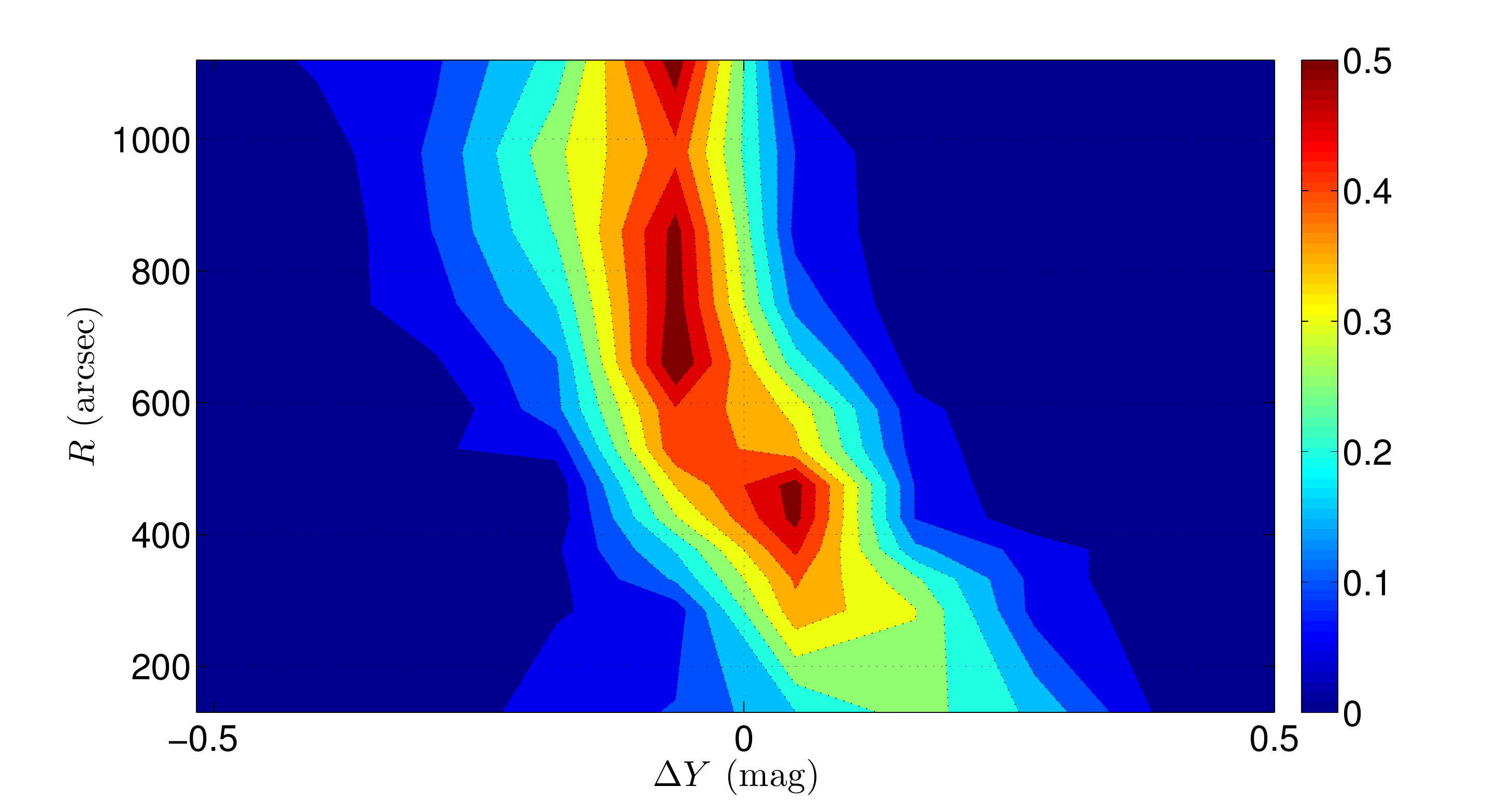}
\caption{2D contours of the SGB stars' probability distribution as a
  function of $\Delta Y$ and radius.}
\label{F8}
\end{figure}

We show the 2D contours of the SGB stars' probability distribution,
$P(\Delta Y,R)$, as a function of $\Delta Y$ and radius in Figure
\ref{F8}. This figure strongly indicates the presence of at least two
SGB populations, the relative fractions of which clearly exhibit a
gradual change in terms of their magnitudes. One population is mainly
concentrated at radii between $R = 600''$ and $R = 900''$, while the
second population peaks between $R = 400''$ and $R = 500''$. It is
possible that a third peak is associated with the innermost cluster
regions, but the significant magnitude dispersion due to contaminating
line-of-sight projection and blending weakens the significance of such
a feature. From Figure \ref{F8}, it is also clear that the typical
magnitudes of the innermost and peripheral SGB stars are characterized
by an offset of $\sim$ 0.2 mag.

\begin{figure}[ht!]
\plotone{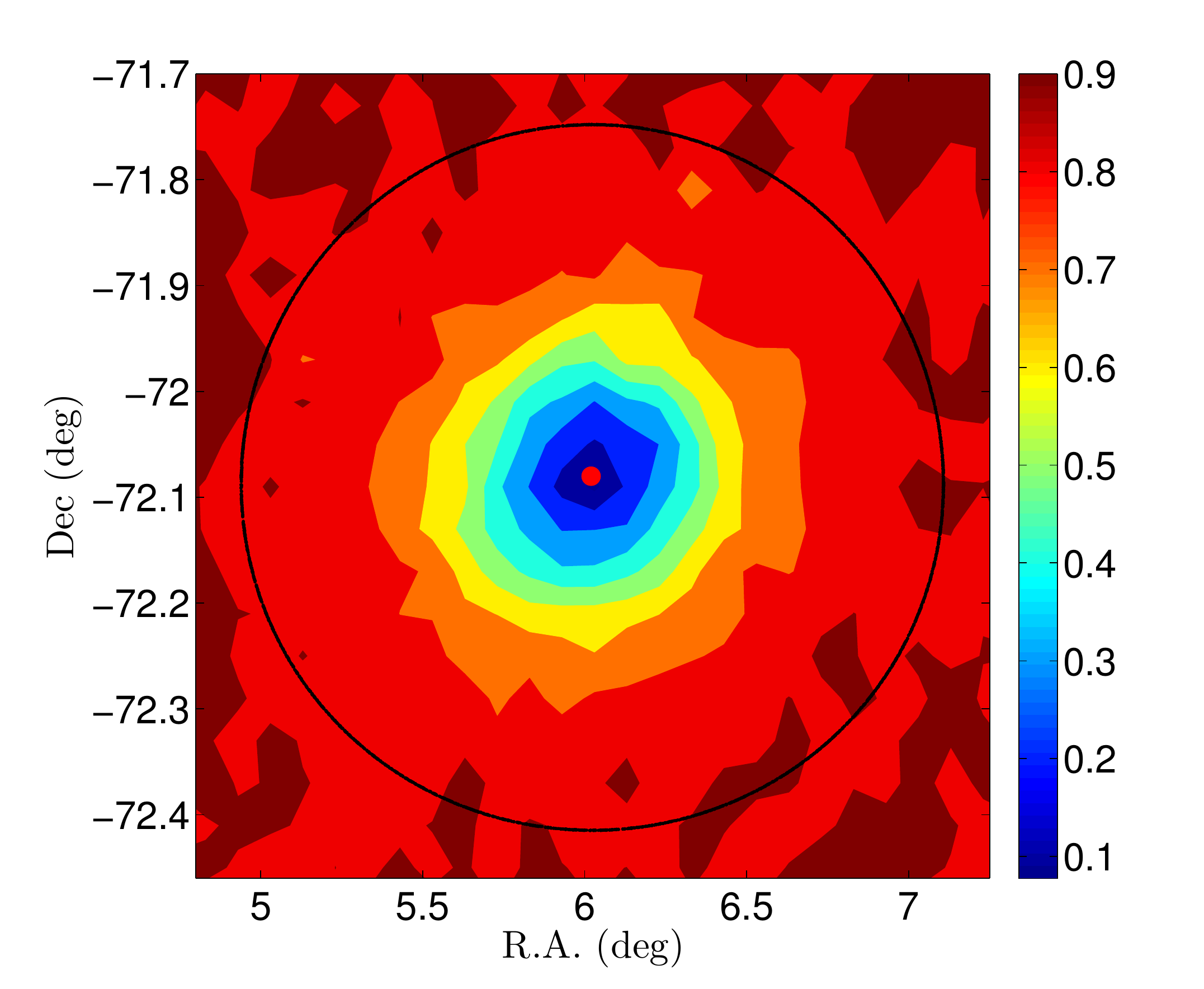}
\caption{Completeness map for SGB stars characterized by $15.4 \le Y
  \le 16.6$ mag. The solid black circle indicates a radius of
  $R=1200''$. The cluster center is shown as a red solid bullet.}
\label{F9}
\end{figure}

Note that SGB stars at $R \la 150''$ cannot be detected because of
crowding in the cluster center; the resulting stellar blends, in
particular of the faint stellar wings of the PSF, cause an increase in
the overall background level. This results in a very small
completeness fraction for the innermost stellar sample, which clearly
increases towards the cluster's periphery. In Figure \ref{F9} we
display the local average completeness for stars with $15.4 \le Y \le
16.6$ mag, corresponding to the magnitude range of the SGB stars
analyzed in this paper. Except for the innermost sample stars, all SGB
stars analyzed here are located in regions where the sampling
completeness is greater than 50\%.

\subsection{Red-giant-branch stars}

\begin{figure}[ht!]
\plotone{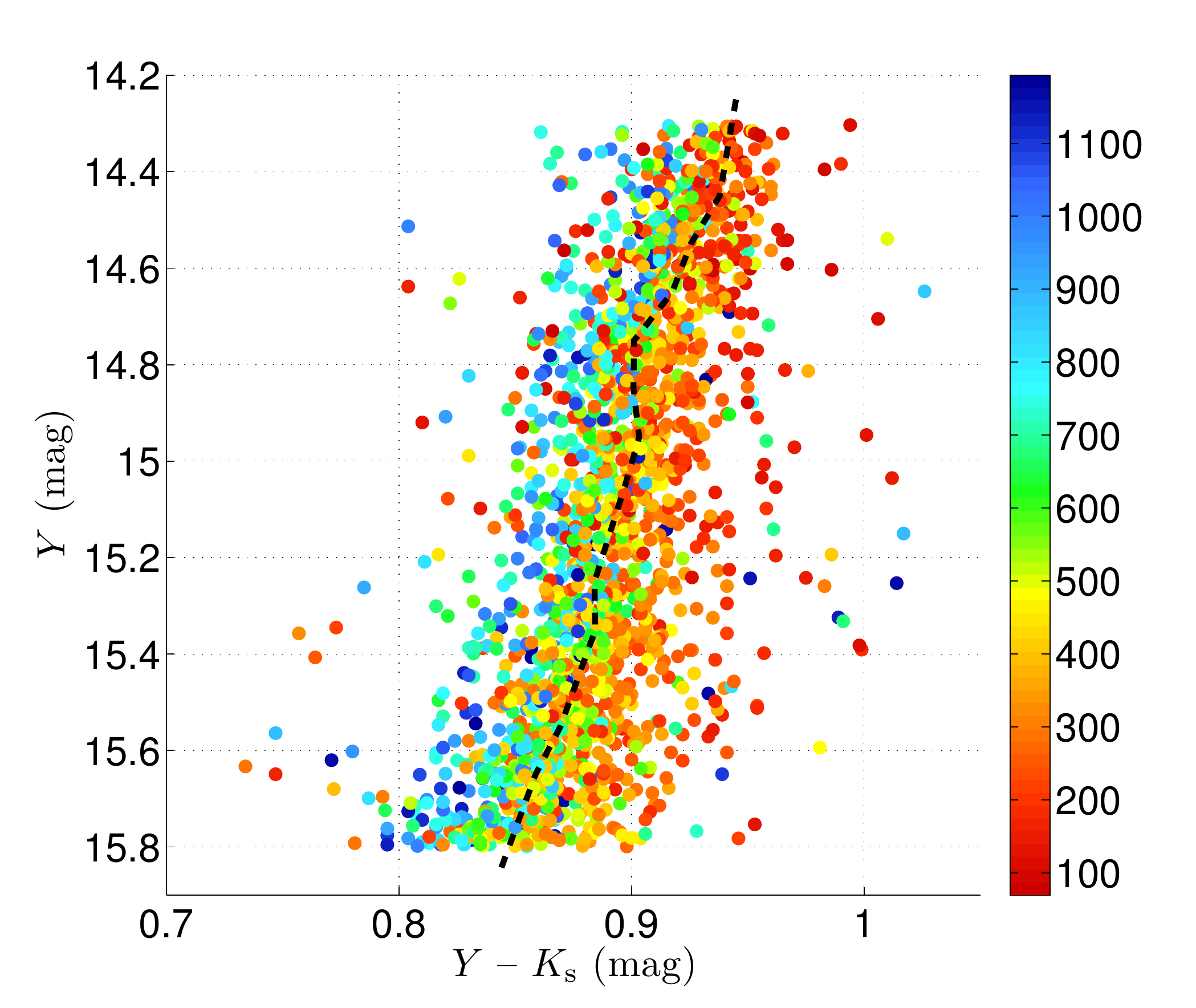}
\caption{CMD of the adopted RGB region, showing only stars found
  within 3$\sigma = 0.135$ mag of the isochrone's ridgeline, where
  $\sigma$ represents the color spread, adopting a Gaussian fitting
  function. The color scale represents distance to the cluster
  center. Because of sampling incompleteness in the cluster core,
  stellar photometry is only robustly available for radii beyond
  $R\sim70''$.}
\label{F10}
\end{figure}

Figure \ref{F3} reveals another striking feature in the form of a
trend between RGB color and radius. Bluer RGB stars are located in the
cluster's outskirts while the innermost RGB stars are clearly redder
and, hence, cooler. This trend is particularly apparent for stars at
the bottom of the RGB. Again, we defined a box that included the
majority of stars at the bottom of the RGB, covering the region
$Y\in[14.3,15.8]$ mag, $(Y-K_{\rm s})\in[0.7,1.2]$ mag. We removed
1352 stars with photometric errors greater than 0.015 mag, which
guarantees that the broadening of the RGB cannot be only caused by
photometric uncertainties. As for the SGB stars, we adopted the
cluster-wide ridge line as benchmark and calculated the color
deviation of the selected RGB sample stars from the corresponding
benchmark color, $\Delta(Y-K_{\rm s})=(Y-K_{\rm s})_{\rm
  RGB}-(Y-K_{\rm s})_{\rm iso}$. We used a Gaussian-like profile to
fit the distribution of the color dispersion, characterized by a
Gaussian $\sigma_{\Delta(Y-K_{\rm s})}=0.045$ mag. We will only
consider the 1850 RGB stars that are found within 3$\sigma = 0.135$
mag of the ridge line, corresponding to 58\% of the full RGB sample in
the area used for our sample selection. Figure \ref{F10} shows the
zoomed-in CMD of the cluster's RGB region.
 
\begin{figure}[ht!]
\plotone{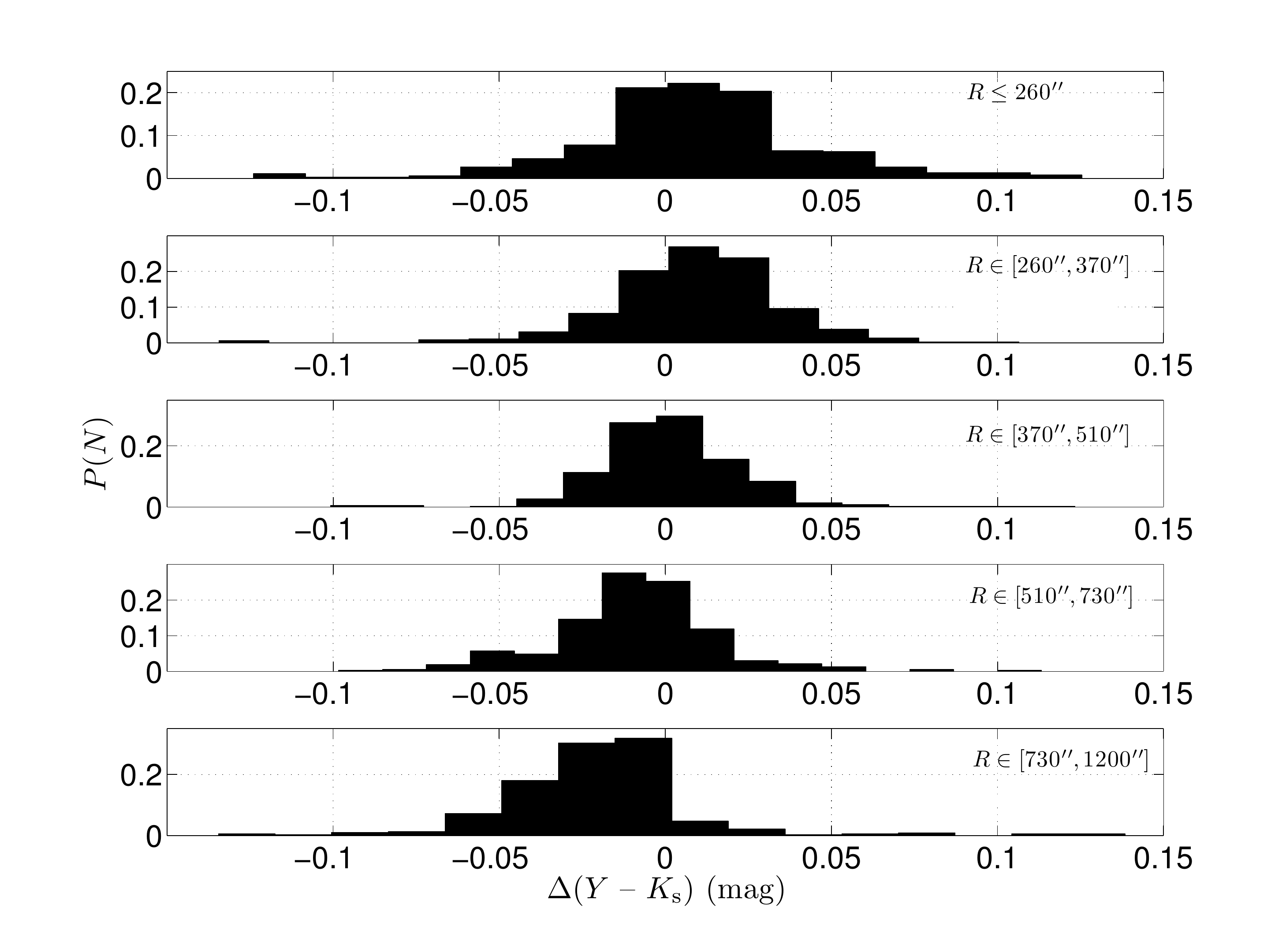}
\caption{Probability distribution of $\Delta(Y - K_{\rm s})$ as a
  function of radius for RGB stars in 47 Tuc. From top to bottom, the
  panels represent radial bins from the cluster's inner regions to its
  outskirts.}
\label{F11}
\end{figure}

We next proceeded with our analysis in a similar manner as for the SGB
stars. We divided the distribution of all RGB stars into five radial
bins, with boundaries at $260''$, $370''$, $510''$, and $730''$, thus
ensuring roughly equivalent numbers of stars (varying from 365 to 377)
in each bin. We then calculated the $\Delta(Y-K_{\rm s})$ probability
distributions (normalized to the total numbers of stars in each bin),
as shown in Figure \ref{F11}. Again, the result shows that the RGB
gradually becomes bluer when going from the cluster's inner regions to
its outskirts, which indicates that the peripheral RGB stars are
systematically hotter than their more centrally located
counterparts. The inner region's RGB, especially for the $R\leq260''$
sample, is quite dispersed compared to the RGB stars at larger
radii. This is again most likely owing to contamination from projected
RGB stars that are physically located at large(r) radii but seen along
more central lines of sight. This blending effect may also cause a
change in the photometric colors of the RGB stars, but because the RGB
is rather vertical, blending will most likely predominantly cause a
shift in magnitude.
 
\begin{figure}[ht!]
\plotone{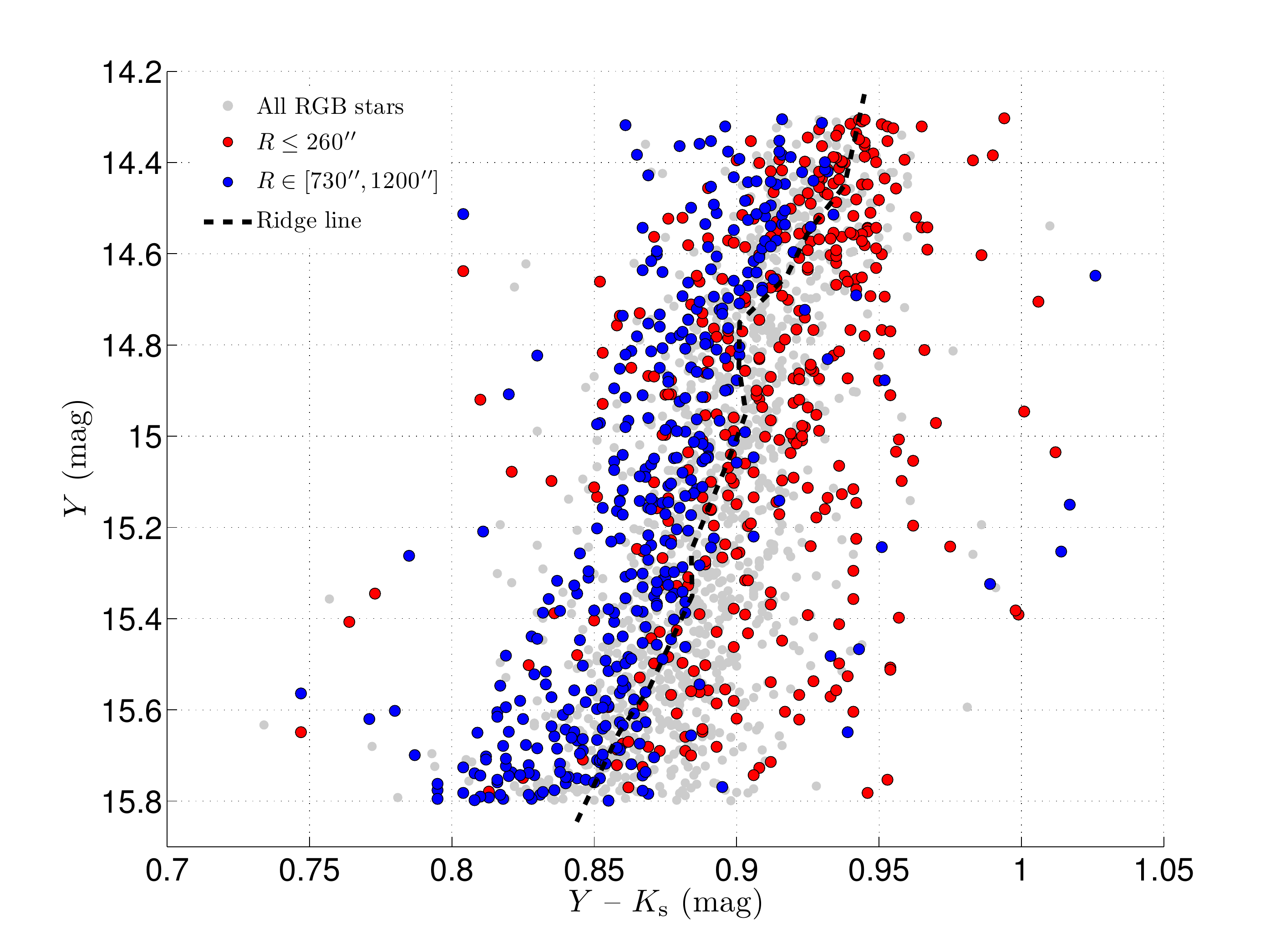}
\caption{CMD comparison of the innermost RGB stars ($R\leq260''$: red
  solid bullets) with the outermost cluster RGB stars
  ($R\in[730'',1200'']$: blue solid bullets). Smaller grey bullets
  represent the full RGB sample.}
\label{F12}
\end{figure}

In Figure \ref{F12} we compare the CMD of the innermost RGB sample
($R\leq260''$) with that drawn from the largest radii associated with
the cluster ($R\in[730'',1200'']$). The behavior of the RGB stars is
similar to that seen for the cluster's SGB stars: both RGB samples
occupy clearly different loci in CMD space, with almost all
peripheral RGB stars being bluer than the benchmark isochrone.

\begin{figure}[ht!]
\plotone{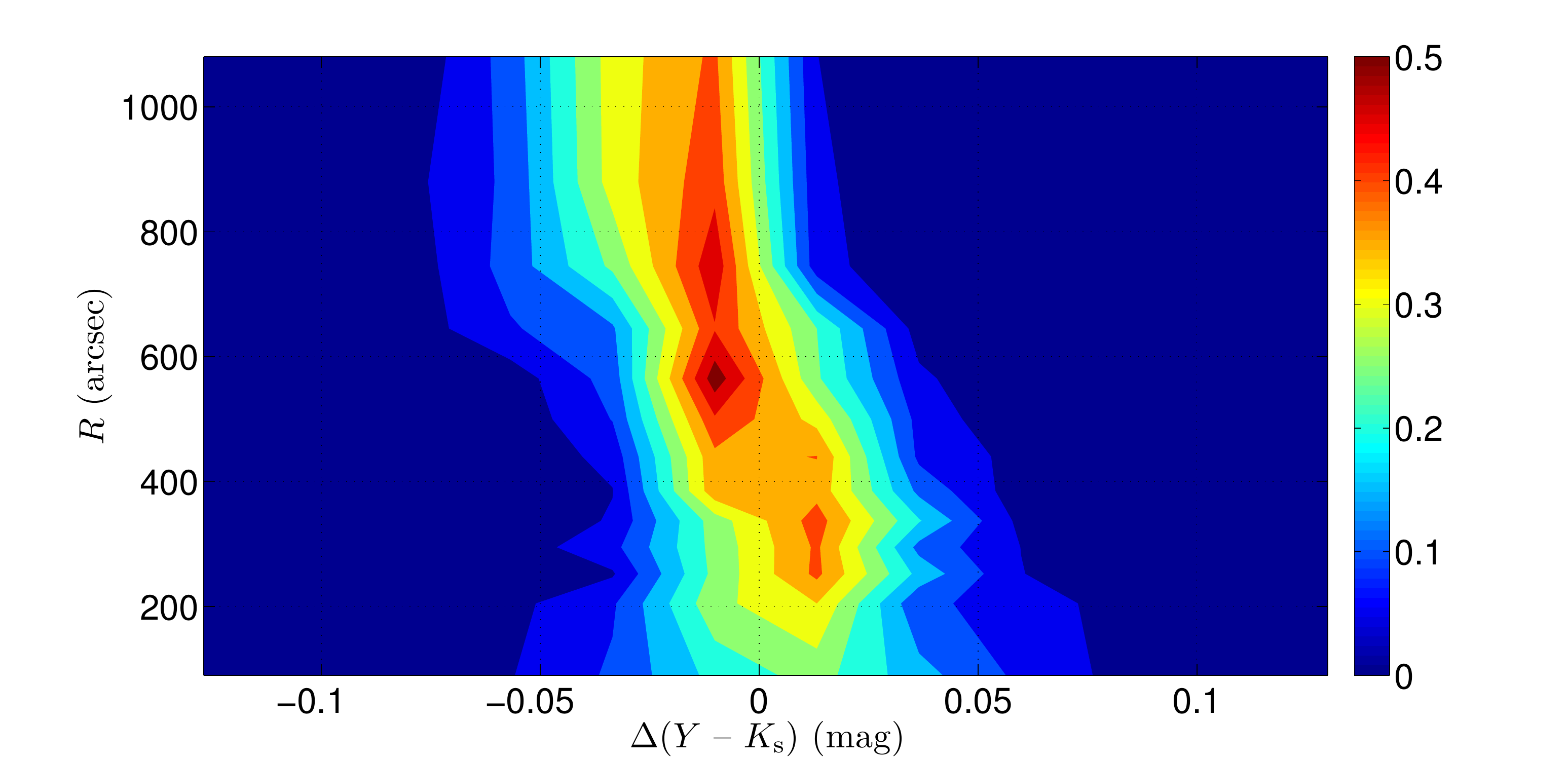}
\caption{2D contours of the RGB stars' probability distribution as a
  function of $\Delta(Y - K_{\rm s})$ and radius.}
\label{F13}
\end{figure}

In Figure \ref{F13} we display the 2D contours of the probability
distribution as a function of both $\Delta(Y-K_{\rm s})$ and
radius. It is clear that the average color and its dispersion
gradually evolve from the cluster's outskirts to the inner
regions. The CMD of the RGB stars located at large radii is narrower
and bluer than that representative of the inner-sample RGB stars. The
probability distribution shows a continuous ridge from the cluster's
outer boundary to $R\sim500''$, followed by a peak between $R=200''$
and $R=400''$ and a significant increase in the color dispersion to
$\geq0.1$ mag in the innermost regions. Such a large color dispersion
cannot be only caused by an intrinsic photometric dispersion, given
that we have constrained the photometric uncertainties (in magnitude)
of all sample stars to $\le 0.015$ mag. The observed width of
$\Delta(Y - K_{\rm s})$ is five times the allowed maximum photometric
broadening due to magnitude errors. Again, we use the Monte Carlo
method to estimate the contamination fraction due to projection for
all radial ranges (see Table 1).

\begin{figure}[ht!]
\plotone{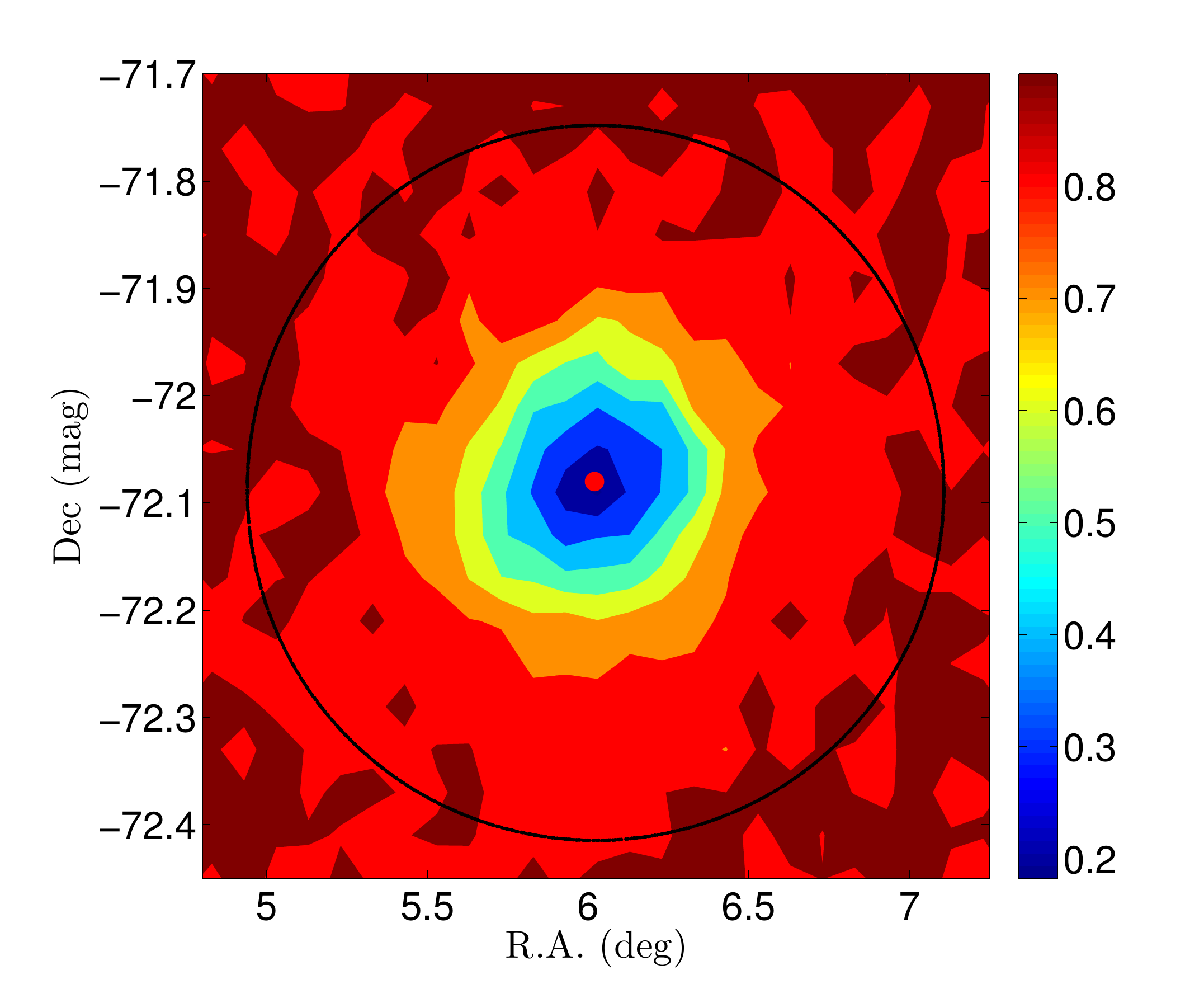}
\caption{Completeness map for RGB stars characterized by $14.2 \le
  Y\le 16.0$ mag. The solid black circle indicates a radius of $R =
  1200''$. The cluster center is shown as a red solid bullet.}
\label{F14}
\end{figure}

Because of the higher levels of sampling incompleteness in the cluster
core, the minimum radius employed in our analysis of the RGB's stellar
population is $R = 70''$. Figure \ref{F14} shows the 2D completeness
map for stars with $14.2 \le Y \le 16.0$ mag. For $R \ga 70''$, all
RGB stars used for our analysis are located in regions where the
completeness fractions are greater than 50\%.

\section{Physical implications}

Before we discuss the possible physical implications of the results
presented in the previous section, we first rule out differential
reddening as the dominant cause of the observed trends for the SGB and
RGB stars in 47 Tuc. We obtained the extinction map of the 47 Tuc area
from the IRAS/DIRBE database provided by the NASA/IPAC Infrared
Science Archive \citep[cf.][]{SFD98,SF11} \citep[see
  also][]{Qing10}. Although the database's native spatial
  resolution is on the order of an arcminute, all details within the
  cluster region have been smoothed: we adopt a radius for the cluster
  region covered by our observations of $20'$, while the radial bins
  used for the analysis are only slightly smaller than $5'$. This
  hence implies that we could potentially resolve any dust
  substructures within the cluster, provided that they exist. Note
  that our reddening corrections may also be applicable to even
  smaller spatial regions, because the spread on the lower RGB is
  larger than that higher up on the RGB; extinction effects cannot be
  used as discriminant in this case. We measured the extinction
distribution from the cluster center to $R = 1200''$ in radial steps
of $5'$. We found that the average extinction in this region varies
from $E(B-V) \simeq 0.027$ mag to $0.028$ mag; the corresponding
extinction at NIR wavelengths is therefore negligible and unimportant
in the context of the observed broadening of both the SGB and RGB in
the 47 Tuc CMD. This result is also consistent with that published by
\cite{Sala07}.

In addition, \cite{Moma12} and \cite{Boye10}
 conclude that there is no evidence that RGB stars in 47
Tuc produce dust. These arguments are supported by the results of
\cite{Bona13}, who map the differential
reddening in 66 GCs (including 47 Tuc; their fig. 4) and conclude that
the main source of differential reddening is interstellar rather than
intracluster dust.

\cite{Milo12c} used helium and nitrogen enhancements to reproduce
  the observed double MS in 47 Tuc, implying that the second
  generation of stars formed from material that had been polluted by
  the first generation. Additional evidence in support of the notion
  that GCs should contain N-rich but CO-depleted second-generation
  stars is found in the GC NGC 6656 (M22), which has been confirmed to
  host two groups of stars characterized by different C+N+O abundances
  \citep{Mari11,Mari12}.  \cite{DiCr10} also suggested that variations
  in C+N+O abundances may be responsible for a broadening of GC SGB
  features, because differences among CNO-group elements will change
  the strength of their compound absorption lines (e.g., OH, NH, CN,
  and CH). Optical filters, particularly those covering wavelengths
  between 3500 {\AA} and 4500 {\AA}, are sensitive to these absorption
  lines \citep[see][their figure 11]{Milo12c}. Variations in CNO
  abundances may hence introduce an additional broadening of CMD
  features observed in the optical wavelength range. However, our data
  have all been obtained in the NIR $Y$ and $K_{\rm s}$ filters, which
  are characterized by effective wavelengths of 1.02 $\mu$m and 2.15
  $\mu$m, respectively. It is, hence, impossible that CNO differences
  would be solely responsible for a significant broadening of the SGB
  and RGB features in the 47 Tuc ($Y, Y-K_{\rm s}$) CMD.

Ejecta from stars of a previous stellar generation will contaminate
the second generation, thus causing $\alpha$ enhancement. However,
previous analyses \citep[cf.][]{Carr04} have not reported any
significant [$\alpha/{\rm Fe}$] dispersion in 47 Tuc. We nevertheless
explored the effects of possible [$\alpha/{\rm Fe}$] variations. We
adopted {\sc pgpuc} isochrones characterized by [$\alpha/{\rm Fe}$] =
0.0 to 0.3 dex for comparison, but found that [$\alpha/{\rm Fe}$]
variations do not significantly change the isochrone's morphology in
either the SGB or RGB phases. Since for these stellar evolutionary
phases the effects of helium and [$\alpha/{\rm Fe}$] differences are
both minor to negligible, we will henceforth adopt ${\rm Y} = 0.26$
and [$\alpha/{\rm Fe}$] = 0.30 dex.

The most straightforward explanation of the observed radial trends in
the magnitudes of the cluster's SGB stars and the colors of its RGB
stars is the presence of multiple stellar populations. Indeed,
\cite{Milo12c} investigated 47 Tuc's apparent triple MSs and found a
significant helium dispersion, ${\rm Y} = 0.256$--0.288 (their table 2
and figure 10, based on the assumption that the presence of a
dispersion in helium content is the only cause of the MS split). This
result is consistent with that of \cite{Nata11}, which is based on
their analysis of RGB-bump stars and HB stars (${\rm
  Y}=0.25$--0.28). Note that these authors did not base their
conclusions on the properties of the cluster's SGB and RGB stars, the
stellar phases we concentrate on here.

 In addition to this variation in the cluster's helium abundance,
  a dispersion in stellar metallicities could also contribute to the
  observed broadening of the 47 Tuc SGB and RGB. A significant number
  of studies have explored this issue, going back to at least
  \cite{Brow92}, who investigated four giant-branch stars in the
  cluster. They found minimum and maximum [Fe/H] abundances of $-0.88$
  dex and $-0.69$ dex, respectively, or $Z$ ranging from $\sim 0.0020$
  to 0.0031. Subsequently, \cite{Carr97} obtained spectroscopic
  observations of three 47 Tuc RGB stars and derived a mean [Fe/H]
  abundance of $-0.70 \pm 0.03$ ($Z$ = 0.0028--0.0033). Given that
  studies based on such small numbers of stars cannot be used to infer
  statistically robust results, we will instead focus on a number of
  more modern studies. \cite{Carr04} analyzed high-dispersion spectra
  of three dwarfs and nine subgiants; they confirmed an [Fe/H] range
  from $-0.59$ dex to $-0.78$ dex ($Z = 0.0026$--0.0041).
  \cite{Koch08} investigated eight RGB and one MSTO stars, and
  determine that the most representative metallicity, [Fe/H] $= -0.76
  \pm 0.05$ dex, which is equivalent to a $Z$ range from 0.0024 to
  0.0030.
  
 We point out that, compared with spectroscopic studies, a number
  of recent articles used photometric measurements to derive a higher
  overall metallicity for the cluster, as well as a larger
  dispersion. \cite{Sala07} used the BaSTI models to analyze their
  photometric observations, yielding $Z = 0.004 \pm 0.001$. (They
  conclude that [Fe/H] $= -0.7 \pm 0.1$ dex, but this determination is
  based on $Z_{\odot} \sim 0.02$; they also find that the
  corresponding $Z = 0.008$, which we suspect is a clerical error.)
  \cite{Nata11} analyzed the 47 Tuc RGB bump stars and concluded that
  they had to assume a model with [M/H] $= -0.50$ or $-0.52$ dex
  (corresponding to [Fe/H] $= -0.53$ or $-0.55$ dex for [M/H] = 0.95
  [Fe/H]) to fit their observations. \cite{Ande09} analyzed the spread
  among the SGB stars. They suggest that if metallicity is the only
  driver of the observed spread, a metallicity dispersion of 0.10 dex
  is implied. These differences compared with spectroscopic
  metallicity determinations may well be owing to an inherent
  selection bias that predominantly affects spectroscopic
  observations. After all, to adequately resolve individual stellar
  spectra, one should try to avoid crowded environments and, hence,
  select candidate stars that are located far from the cluster
  center. For instance, \cite{Carr04} selected their sample stars at
  distances in excess of $800''$ from the 47 Tuc core. To reconcile
  the differences between the photometric and spectroscopic
  metallicity determinations in the context of the analysis presented
  in this article, we adopt $Z = 0.0041$ from \cite{Carr04} as the
  typical 47 Tuc metallicity, but we assume that the dispersion in
  metallicity is represented by that implied by the photometric
  analyses, i.e., $\Delta_{\rm [Fe/H]} \sim 0.10$ dex (corresponding
  to a full metallicity range given by $Z = 0.0033$--0.0051).

\begin{figure}[ht!]
\plotone{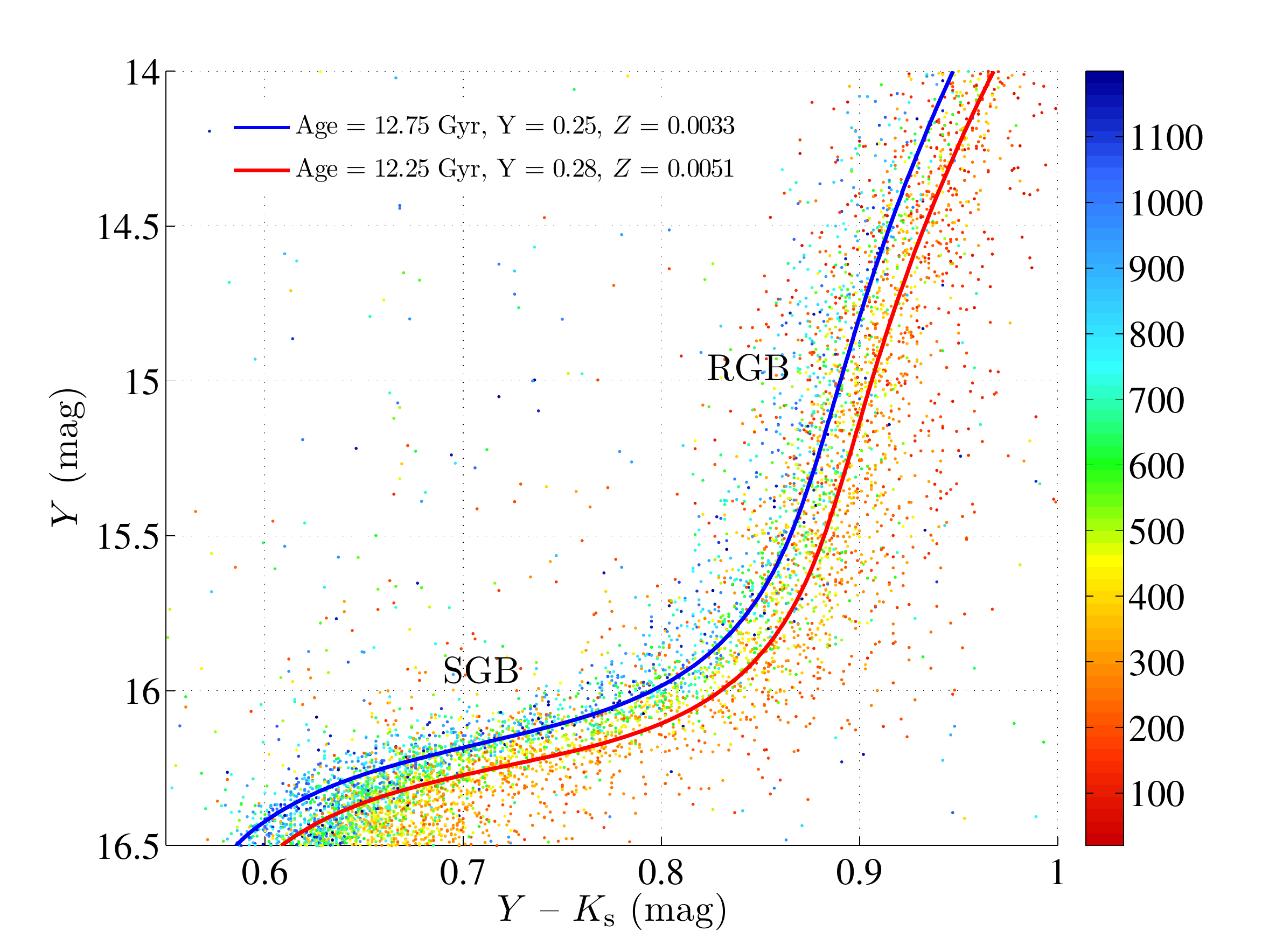}
\caption{Isochrone fits of a young, helium-rich, and metal-rich model
  (red line), as well as of an old, helium-poor, and metal-poor model
  (blue line). The color scale represents the stars' distances to the
  adopted cluster center (in arcsec).}
\label{F15}
\end{figure}

 We used a Monte Carlo method to mimic the distribution in CMD
  space of the observed SGB and RGB stars, based on adoption of two
  suitable `bracketing' isochrones. We follow the usual assumption
  that the helium-rich generation of stars should also be metallicity
  enhanced and adopt the highest-dispersion metallicity range from the
  literature for our model isochrones, i.e., $Z$ = 0.0033--0.0051 and
  Y = 0.25--0.28. The helium and metallicity abundances must be varied
  only between these values. To generate a physically realistic
  broadening of the CMD, we also considered the possible presence of
  an age dispersion among the cluster's member stars. At the age of 47
  Tuc, even a very small age dispersion represents a long time span,
  which may be enough for the brightest first-generation stars to
  evolve to late evolutionary stages. Previous studies have suggested
  a possible age dispersion among 47 Tuc stars of order 1 Gyr
  \citep[$\sim$12--13 Gyr; see][]{DiCr10}. \cite{Grat03} found a
  model-dependent best-fitting age for 47 Tuc of either $11.2\pm1.1$
  Gyr or $10.8\pm1.1$ Gyr, while \cite{Zocc01} determined an age of
  $13\pm2.5$ Gyr. Some recent papers have reported smaller possible
  age dispersions, with a typical age span of $12.8 \pm 0.6$ Gyr
  \citep{Mari09}, $12.75 \pm 0.5$ Gyr \citep{Dott10}, or even $11.75
  \pm 0.25$ Gyr \citep{Vand13}. We hence conservatively select a
  minimum age difference of 0.25 Gyr for our model, e.g., $12.50 \pm
  0.25$ Gyr. We thus generate an artificial 47 Tuc CMD defined by Y =
  0.25--0.28, $Z$ = 0.0033--0.0051, and an age range spanning from
  12.25 Gyr to 12.75 Gyr. In Figure \ref{F15} we display the
  corresponding range in best-fitting isochrones.

\begin{figure}[ht!]
\plotone{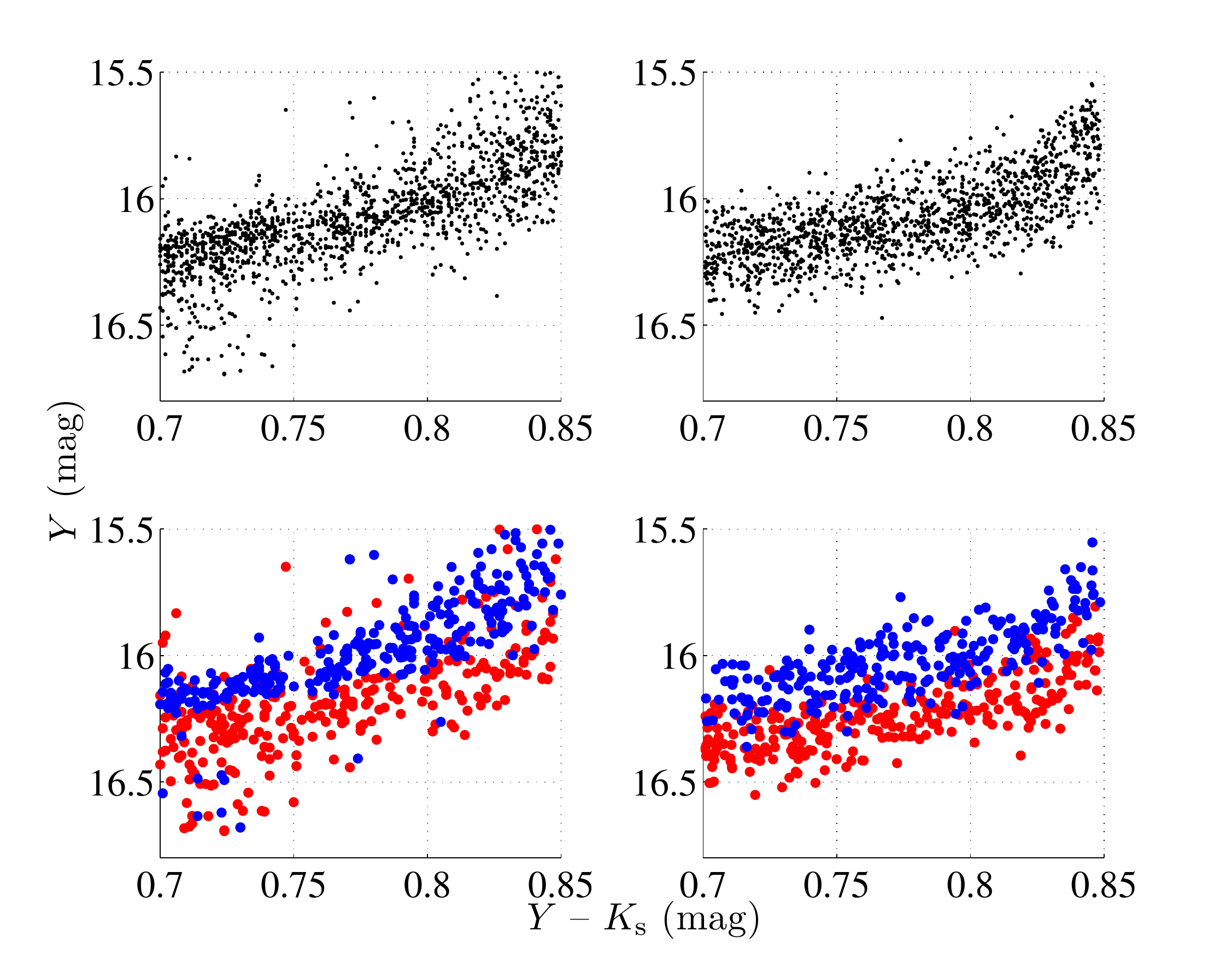}
\caption{Top left: Observed overall distribution of SGB stars. Top
  right: Simulated SGB stars. Bottom left: Observed innermost SGB
  stars ($R\leq340''$: blue solid bullets) and outermost SGB stars
  ($R\in[730'', 1200'']$: red slid bullets). Bottom right: Simulated
  innermost SGB and outermost SGB stars.}
\label{F16}
\end{figure}

\begin{figure}[ht!]
\plotone{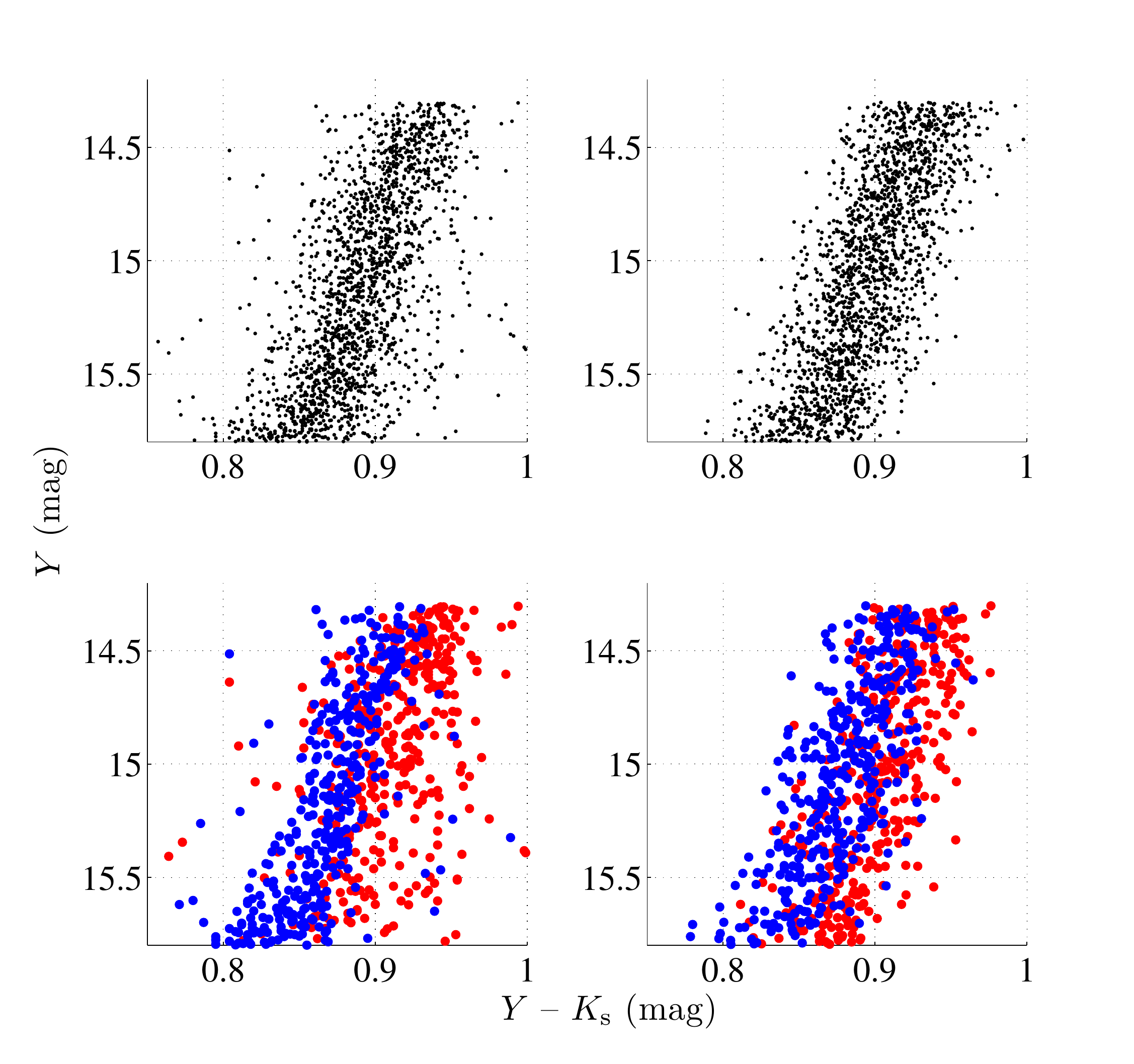}
\caption{As Figure \ref{F16}, but for RGB stars. The innermost sample
  contains stars with $R\leq260''$, while the outermost sample covers
  $R\in[800'',1200'']$.}
\label{F17}
\end{figure}

If we only mimic stars with photometric errors of up to 0.015
  mag, we cannot reproduce the observations. Once again, this confirms
  that the observed SGB and RGB broadening is more likely caused by a
  dispersion in stellar generations and not solely by photometric
  uncertainties. Note that the typical 1$\sigma_{\Delta{Y}}$ and
  1$\sigma_{\Delta{(Y-K_{\rm s})}}$ dispersions of the SGB and RGB
  stars are 0.18 mag and 0.045 mag, respectively (see the captions of
  Figures \ref{F5} and \ref{F10}). The large, observed dispersion is
  owing to a combination of photometric uncertainties and a dispersion
  in stellar-generation properties (age, helium abundance, and
  metallicity). However, to avoid complications associated with
  adopting a full age range, we still assume that the bulk of the
  cluster stars can be adequately described by two separate
  generations of stars (but see below), although we relax their
  dispersions to cover 0.5$\sigma_{\Delta{Y}} = 0.090$ mag and
  0.5$\sigma_{\Delta{(Y-K_{\rm s})}} = 0.022$ mag for the simulated
  SGB and RGB stars, respectively.  This still reproduces two apparent
  branches in the CMD. In Figures \ref{F16} and \ref{F17} we display
  the simulated CMDs for the SGB and RGB stars (right-hand panels),
  compared with the real observations (left-hand panels). The top two
  panels represent the overall comparison, while the bottom two panels
  only indicate the observed inner- and outermost samples, as well as
  their simulated counterparts. The simulated SGB and RGB samples
  contain the same numbers of stars as the observations. We adopt a
  flat magnitude distribution, because the magnitude range covered by
  the SGB stars is rather narrow, while for the RGB stars we are only
  concerned with their color dispersion. Hence, any difference in the
  luminosity function will not affect our RGB-based analysis. We also
  confirmed that the observed luminosity function pertaining to our
  selected sample of RGB stars is close to flat.

 Instead of checking the performance of our fits by eye, we
  quantify the similarity between the observed and simulated CMDs
  using $\chi^2$ minimization. At a given radius, we assume that the
  simulated CMD contains a certain fraction of first-generation stars,
  $f_{\rm fg}$. The remainder of the stellar population at that radius
  then follows the second-generation isochrone, which hence is
  characterized by a fraction $f_{\rm sg} = 1 - f_{\rm fg}$. We then
  use the adopted ridge line (see Figures \ref{F5} and \ref{F10}) to
  calculate the probability distributions of $\Delta{Y}$ for the
  simulated SGB stars, and of $\Delta (Y - K_{\rm s})$ for the
  simulated RGB stars, similarly to what we did for the observations
  in Figures \ref{F6} and \ref{F11}. As a function of radius, we then
  calculate the corresponding $\chi^2$ value,
\begin{equation}
   \chi^2=\sum_{n}\frac{({N}'-{N})^2}{N},
\end{equation}
where $N'$ ($N$) indicates the simulated (observed) number of stars in
different $\Delta{Y}$ or $\Delta (Y - K_{\rm s})$ bins (for SGB and
RGB stars, respectively), and $n$ represents the number of bins. The
$\chi^2$ value indicates the level of similarity between the simulated
and observed CMDs. We vary $f_{\rm{fg}}$ from 5\% to 95\% and
determine the minimum $\chi^2$ value for each input $f_{\rm fg}$. We
find that the use of a parabolic function can approximate the
$\chi^2(f_{\rm{fg}})$ distribution very well: it will yield both the
global minimum $\chi^2$ value (and, hence, the best-fitting $f_{\rm
  fg}$) and its 1$\sigma$ uncertainty \citep[see also][]{grijs13}. The
latter corresponds to the difference between $\chi^2_{\rm{min}}$ and
$\chi^2_{\rm{min}}+1$ \citep{Avni76,Wall96}. We repeat each
calculation 100 times and record the average value as the
corresponding $\chi^2$. We thus eliminate random scatter in the
$\chi^2$ values and obtain a smooth $\chi^2$ curve as a function of
$f_{\rm fg}$.

\begin{figure}[htb!]
\plotone{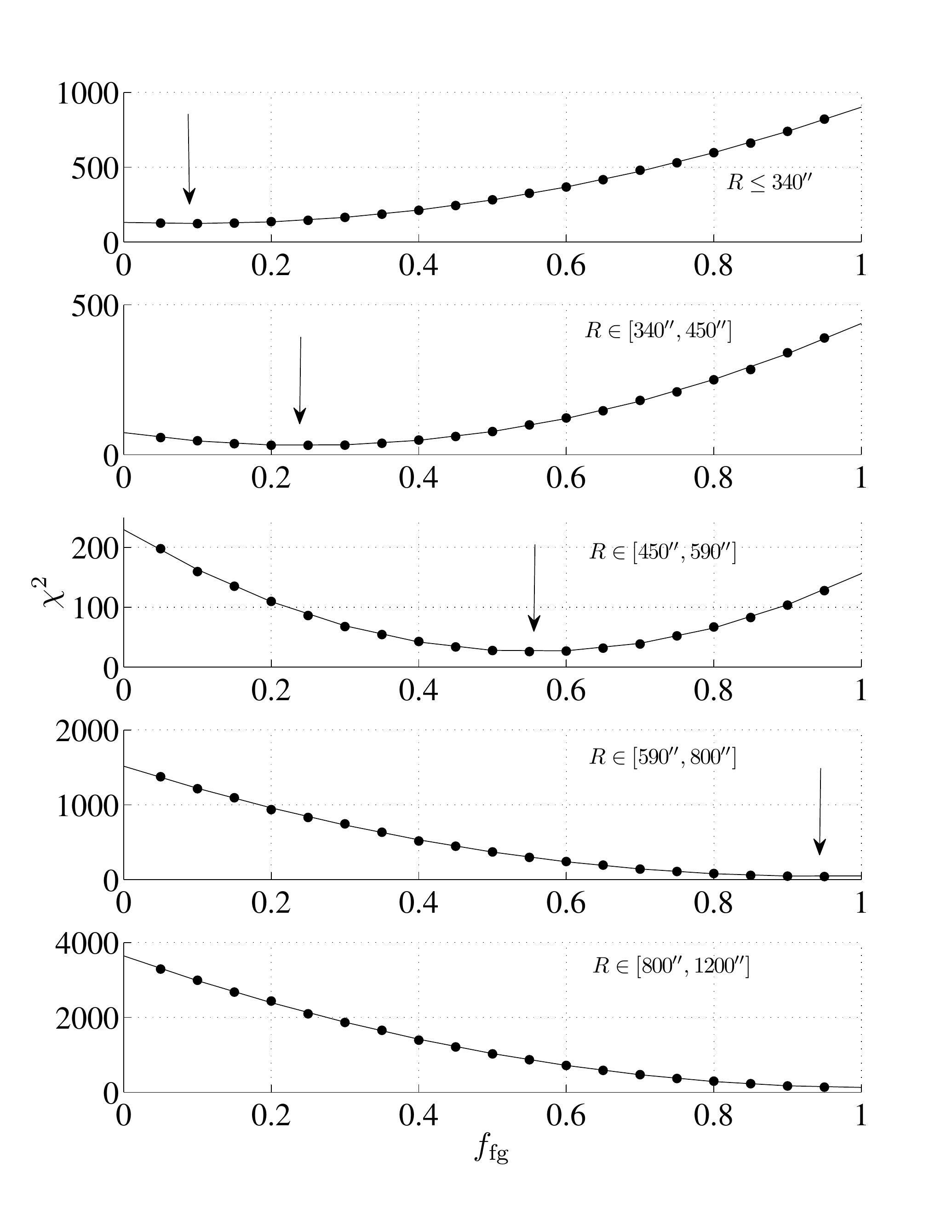}
\caption{$\chi^2$ as a function of $f_{\rm fg}$ for the SGB
  stars. From top to bottom, the panels indicate the results from the
  innermost region to the cluster outskirts. The curve represents the
  best-fitting parabolic function; the arrows indicate the minimum
  $\chi^2$ values and, hence, the best-fitting $f_{\rm fg}$. The
  bottom panel exhibits a best-fitting $f_{\rm fg}$ of unity.}
\label{F18}
\end{figure}

\begin{figure}[htb!]
\plotone{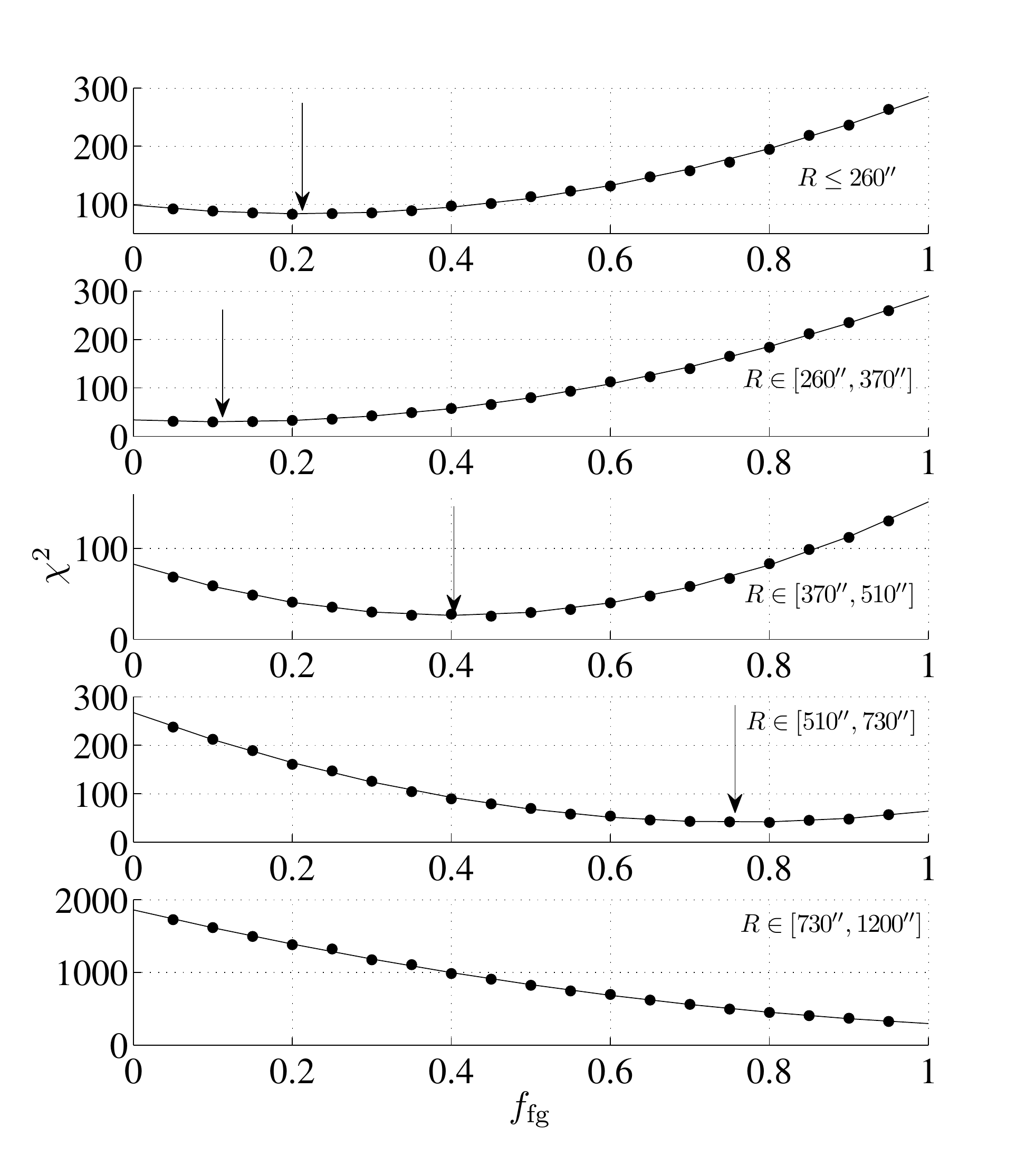}
\caption{As Figure \ref{F18}, but for the RGB stars.}
\label{F19}
\end{figure}

In Figures \ref{F18} and \ref{F19}, we show the calculated $\chi^2$
distributions for different input fractions of $f_{\rm fg}$, as well
as the best-fitting parabolic curves, for the SGB and RGB samples,
respectively. From top to bottom, the panels display the results for
the innermost to outermost radial samples; the arrows represent the
best-fitting $f_{\rm fg}$ fractions. For both the SGB and RGB stars,
the best-fitting $f_{\rm fg}$ fractions for the outermost samples are
100\%. This indicate that at those radii, the composition of 47 Tuc is
still close to that expected for a simple stellar population. It is
also clear that for both the SGB and RGB stars, the first-generation
stars only occupy a small fraction in the innermost region, but that
this fraction increases significantly toward the cluster's outskirts,
reaching unity at the adopted outer boundary, $R = 1200''$.

\begin{figure}[htb!]
\plotone{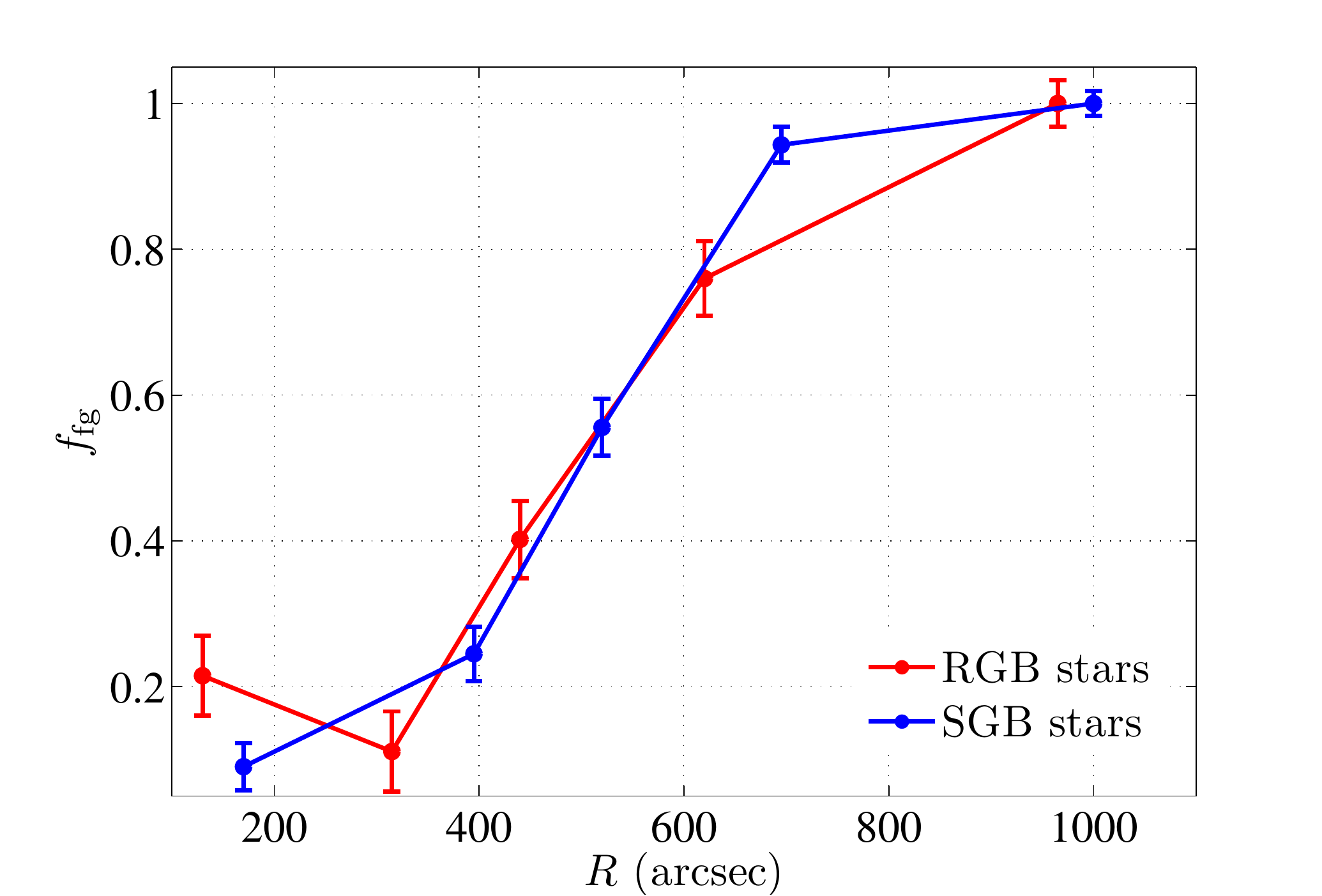}
\caption{Best-fitting $f_{\rm fg}$ fraction as a function of radius
  for SGB stars (blue) and RGB stars (red). The error bars represent
  the 1$\sigma$ statistical uncertainties, which are all less than
  5.5\%.}
\label{F20}
\end{figure}

In Figure \ref{F20} we display the best-fitting $f_{\rm fg}$ fraction
as a function of radius, for both the SGB stars (blue line) and their
RGB counterparts (red line). The error bars indicate the 1$\sigma$
statistical uncertainties, which are all less than 5.5\%. The
innermost sample is clearly only composed of a very small fraction of
first-generation stars, which implies that most member stars should
belong to the second generation. On the other hand, it appears that
the outermost region can still be well described by a simple stellar
population. In addition, even if we investigate the SGB and RGB stars
separately, both exhibit consistent trends in $f_{\rm fg}$ from $R
\sim 400''$ to $R \sim 600''$. This strongly implies that they form
intrinsically uniform samples: both their inner- and outermost members
share a continuous track in the CMD. However, compared with the SGB
stars in the innermost region, the RGB stars in the same region show a
relatively large $f_{\rm fg}$ fraction, which may be caused by the
large projected contamination and small completeness in this
region. We summarize these statements quantitatively in Table 2.

\begin{table}[h!]
 \centering
 \begin{minipage}{110mm}
  \caption{Best-fitting $f_{\rm fg}$ fractions as a function of
    radius.}
  \begin{tabular}{@{}l|c||l|c@{}}
  \hline
  \hline
SGB stars           & $f_{\rm fg}$ & RGB stars   & $f_{\rm fg}$ \\ 
\hline\hline
$R\leq340''$        & 9.02\% & $R\leq260''$        & 21.50\% \\
$R\in[340'',450'']$ & 24.50\% & $R\in[260'',370'']$ & 11.10\% \\
$R\in[450'',590'']$ & 55.57\% & $R\in[370'',510'']$ & 40.20\% \\
$R\in[590'',800'']$ & 94.31\% & $R\in[510'',730'']$ & 76.00\% \\
$R\in[800'',1200'']$ & 100\% & $R\in[730'',1200'']$ & 100\% \\
\hline\hline
\end{tabular}
\end{minipage}
\end{table}

In summary, the observed distributions of SGB and RGB stars in the 47
Tuc CMD can best be explained if the cluster's SGB and RGB stars in
the outer regions are both helium- and metal-poor, while their more
centrally located counterparts are helium- and metal-rich.

Our results decisively confirm the presence of second-generation stars
in 47 Tuc. \citet[][their figure 33]{Milo12c} calculated the number
frequency of red RGB stars (which they refer to as ``RGBb''), which
increases from the cluster's periphery (RGBb/RGB $\sim 60$\%) to its
central region (RGBb/RGB $\sim 80$\%), while the average RGB color and
its dispersion remain roughly constant. A similar result is presented
by \cite{Cord14}, who found that the fraction of RGBb stars increases
from $\sim$40\% to $\sim$90\% from the cluster's outskirts to its
central regions. Our analysis yields a fraction of 5--10\% in the
  outermost region to $\sim$90\% in the cluster core. This is a
similar although even more significant trend as that reported by
\citet{Milo12c}. Our result is also consistent with the $N$-body
simulations of \cite{Vesp13}.
  
This scenario is supported by the spatial distributions of the SGB and
RGB stars. The second generation stars would more likely have formed
within the denser core region of the cluster, and their clearly
different distributions suggest that the second-generation stars may
originally have been dominated (in terms of stellar numbers) by
first-generation stars, but that most first-generation stars
subsequently somehow escaped from the cluster. This may be owing to
expansion of the cluster triggered by either Type II supernovae
\citep{Decr08,Derc08} or tidal stripping \citep{Milo12c}, which
eventually led to a system with second-generation stars preferentially
distributed closer to the cluster center.

The analysis presented in this article is the first investigation of
the radial behavior of the multiple populations of SGB stars in 47
Tuc. The suggested helium-abundance dispersion of $\Delta{Y}=0.03$
fully concurs with that derived by \cite{Ande09},
$\Delta{Y}=0.026$. Figure \ref{F8} shows that this helium enhancement
may be most significant near $R\sim500''$, where one can discern two
clear peaks on either side of this radius. Our simulation results
  also confirm this result. For both SGB and RGB stars, the $f_{\rm
    fg}$ fractions increase significantly from $R \sim 400''$ to $R
  \sim 600''$, exactly covering the locations of these two peaks. This
  indicates that pollution by a second generation of stars becomes
  apparent at these radii. The discovery of a population of more
centrally concentrated helium-rich stars is also consistent with the
analysis of \cite{Bekk11}, \cite{Nata11}, and \cite{Vent14}. The
derived enhanced metallicity implies an origin related to the ejecta
of massive stars \citep{Decr07,Mink09,Vent09,Cord14}.

\section{Conclusions}

We have presented a deep, wide-area, NIR CMD of---and the
corresponding spatial distribution of the SGB and RGB stars in---47
Tuc, obtained with the VISTA telescope. Apparent differences are seen
between SGB and RGB stars in the cluster's outskirts compared with its
inner regions. The peripheral SGB stars are systematically brighter
and define a narrower CMD feature than the SGB stars in the cluster
core. RGB stars in the cluster's outskirts are bluer than the
innermost RGB stars.

We adopted the isochrone defined by the cluster's overall physical
properties as our benchmark and carefully investigated the magnitude
spread of the SGB stars as a function of radius. The resulting 2D
probability distribution displays two clear peaks between $R = 600''$
and $R = 900''$, and between $R = 400''$ and $R = 500''$. We followed
a similar approach in our analysis of the cluster's RGB stars. The
corresponding probability distribution shows a continuous ridge from
the cluster's outer edge to $R \sim 500''$, followed by a peak between
$R = 200''$ and $R = 400''$. For both SGB stars and RGB stars, the
stars located at the largest radii are clearly different from
the innermost sample stars.

We used a Monte Carlo method to estimate the contamination due to
line-of-sight projection as a function of radius and conclude that the
innermost radial ranges are, in fact, seriously contaminated by
projected peripheral stars. This explains the larger magnitude (color)
dispersion of the innermost SGB (RGB) stars compared with their
counterparts at larger radii. In fact, if we could properly
reconstruct the 3D distribution of the observed stars, the inner
stellar sample would most likely be more clearly different from the
outer sample.

We use different models in our attempts at explaining the
observations, including dispersions in helium abundance, [$\alpha/{\rm
    Fe}$], metallicity, and age. The most promising explanation is
that the observed SGB and RGB stars are characterized by variations in
helium and metal abundance, with the peripheral stars being both more
helium-poor (Y = 0.25) and more metal-poor ($Z = 0.0033$), while the
more centrally located stars are likely both more helium-rich (Y =
0.28) and more metal-rich ($Z = 0.0051$). The effects of [$\alpha/{\rm
    Fe}$] variations are negligible; our photometric data are
insufficiently sensitive to trace any such differences. The
  helium-rich, metal-rich sample is also consistent with a relatively
  younger isochrone, with age of 12.25 Gyr, while the helium-poor,
  metal-poor sample follows a 12.75 Gyr isochrone. An age dispersion
of $\sim 0.5$ Gyr is required to best match the morphologies of both
the SGB and RGB stars. In this context, the helium-abundance and
metallicity dispersion invoked to explain the majority of the
broadening of the CMD features would have originated from
contamination by first-generation stars. We generate a series of
  simulated CMDs covering both the SGB and RGB stars, and use a
  $\chi^2$-minimization method to quantify the best-fitting
  first-generation stellar fraction. The result shows a clear,
  increasing trend from the innermost region to the cluster outskirts.
  This indicates that only a very small fraction of first-generation
  stars is contained in the cluster core, while the stellar population
  in the outskirts is close to a simple stellar population. The
  $\chi^2$-minimization results also show excellent agreement for both
  the SGB and RGB stars, both as regards the trend and the actual
  values, at least for $f_{\rm fg}$ at $R\in[400'', 600'']$, which
  strongly indicates that the SGB and RGB stars share the same stellar
  population composition.
  
Based on the analysis presented in this article, we thus confirm that
47 Tuc is composed of more than one stellar population. The most
straightforward interpretation of the origin of a second stellar
generation is that it is the remnant of the first stellar generation
with enhanced helium abundance (from Y = 0.25 to Y = 0.28), as well as
enhanced metallicity (from $Z = 0.0033$ to $Z = 0.0051$).

\section*{Acknowledgements}

The analysis in this article is based on observations made with the
VISTA telescope at the European Southern Observatory under program ID
179.B-2003. We thank the team responsible for the UK's VISTA Data Flow
System---comprising the VISTA pipeline at the Cambridge Astronomy
Survey Unit (CASU) and the VISTA Science Archive at the Wide Field
Astronomy Unit (Edinburgh; WFAU)---for providing calibrated data
products, supported by the UK's Science and Technology Facilities
Council. Partial financial support for this work was provided by the
National Natural Science Foundation of China (NSFC) through grants
11073001 and 11373010 (RdG, CL). RG is a Pegasus postdoctoral fellow
supported by the Fonds Wetenschappelijk Onderzoek (FWO), Flanders.
B-QF is the recipient of a John Stocker Postdoctoral Fellowship from
the Science and Industry Research Fund (Australia). AEP acknowledges
partial support from CONICET (Argentina) and the Agencia Nacional de
Promoci\'on Cient\'{\i}fica y Tecnol\'ogica (ANPCyT).

\end{document}